# Quantum Scattering Theory in the light of an exactly solvable model with rearrangement collisions

S. Varma* and E. C. G. Sudarshan †

Center for Particle Physics and Department of Physics,
The University of Texas at Austin, Austin, Texas 78712

ABSTRACT

We present an exactly solvable quantum field theory which allows rearrangement collisions. We solve the model in the relevant sectors and demonstrate the orthonormality and completeness of the solutions, and construct the S-matrix. In the light of the exact solutions constructed, we discuss various issues and assumptions in quantum scattering theory, including the isometry of the Möller wave matrix, the normalization and completeness of asymptotic states, and the non-orthogonality of basis states. We show that these common assertions do not obtain in this model. We suggest a general formalism for scattering theory which overcomes these, and other, shortcomings and limitations of the existing formalisms in the literature.

## 1. Introduction

Quantum scattering has been an important subject of study since the early days of quantum physics. Unfortunately, while we have a reasonable understanding and intuition for simple scattering problems, such as single channel scattering, we cannot say the same for more general scattering problems such as multi-channel scattering, rearrangement collisions, field theoretic scattering, problems where bound states appear, and the like. There have been many attempts to generalize scattering theory to deal with more complicated cases. However, the literature in this field, though vast, is highly implicit and not constructive. Most authors that have dealt with the problem have carried over the intuition developed from the study of single channel potential scattering. This intuition, while quite adequate for simple problems, is ill-equipped to deal with more complicated scattering problems. Therefore, it is important to examine the common claim by some authors, for example Haag [1,2], that their formalism is general enough to encompass complicated scattering problems, as well as field theory. Unfortunately, most such formalisms are based largely on previous results from potential scattering. Furthermore, even when these problems are addressed in quantum mechanical scattering, field theoretic scattering remains problematical. Many papers, such as the paper by Gell-Mann and Goldberger [3], treat field theoretic scattering as somewhat of an afterthought, without much development from first principles, or such as the papers by Van Hove [4], treat it as a case for discussion. The first clear development of field theoretic scattering from first principles was the seminal paper by Lehmann, Symanzik, and Zimmerman [5]. However the LSZ formalism is not applicable in many cases, for example, collisions in which stable bound states appear. This is, in fact, pointed out by the authors themselves.

* e-mail: varmint@delphi.com

† e-mail: sudarshan@physics.utexas.edu

All this leads to the question: how many of our results and assumptions, and how much of our intuition can we carry over from simple single-channel potential scattering to more complicated scattering situations? To attempt to answer this question, we will construct an exactly solvable sector for a quantum field theory. This model has a three particle sector, and allows rearrangement collisions. We will use the solutions of this model, along with previous results, to point out where the existing formalism has defects and shortcomings.

Our model, which we shall call the Rearrangement Model, is an elaboration of the Lee Model [6], and the Cascade Model [7], but with extra particles and couplings chosen in such a way as to allow rearrangement collisions. The couplings of the model are $B \leftrightarrow C\phi$ and $D \leftrightarrow C\theta$. This model has a sector, which we shall call the Rearrangement Sector, $B\theta \leftrightarrow C\theta\phi \leftrightarrow D\phi$, in which rearrangement collisions can take place. The model can be applied directly to physical problems involving rearrangement collisions. We shall, however, leave the applications to subsequent work.

This model is interesting because it a very simple one, and yet contains the essence of many phenomena that can take place in an interacting system. It displays the following characteristics:

1. New states can appear, which have no corresponding states in the original Hamiltonian.
2. The thresholds and continuous spectra shift, and the spectra of $H$ and $H_0$ are *not* the same. Furthermore, the continuous spectra are shifted by different amounts.
3. Genuine rearrangement collisions can take place. Yet we have the sub-additivity of the spectra: the spectra in the higher sectors is the sum of all the spectra in the lower sectors, with possibly additional terms.

We will construct the solutions of this model, and then, in the light of the solutions we have constructed, will examine various assumptions and assertions made in the literature about quantum scattering theory. In particular, we will focus on four key points, that of the isometry [1] of the Möller wave matrix [8], the normalization [1] and completeness [9] of the asymptotic states, and the non-orthogonality of the physical $B\theta$, $C\theta\phi$, and $D\phi$ states [10,11]. We shall show that these assertions do not obtain in this model. We also comment upon the use of the (renormalized) free Hamiltonian in the literature [1,3,9,10,12], rather than the correct prescription, which is to use the isospectral comparison Hamiltonian (see section 8).

The plan of this work is as follows. We start with a review of scattering theory in both the single channel and multiple channel cases. In the next three sections, we introduce the Hamiltonian of the Rearrangement Model, show how explicit solutions can be found for this model in the Rearrangement Sector, and verify that the solutions obtained are, in fact, solutions to our model. We then show that the solutions obtained are orthonormal and complete, write down the Möller matrix and the comparison Hamiltonian, and show that the comparison Hamiltonian is isospectral with the full Hamiltonian, but not with the free Hamiltonian. Then we calculate the S-matrix of the system in this sector, demonstrate its unitarity, and calculate its eigenphases. In section 12, we present a general formalism for scattering theory which overcomes the shortcomings and limitations of the existing formalisms in the literature. In section 13 we discuss scattering theory and its relation to the solutions that we constructed, and to previous work. Finally, in section 14, we summarize our work and present our conclusions.

## 2. The Single Channel Formalism

For our purposes, it makes no difference whether we use the time dependent or time independent formalisms of scattering theory. We are concerned with the assumptions and results that obtain from the formalisms, and they remain essentially the same in both cases. We shall therefore restrict ourselves to the time dependent formalism in the next two sections. We follow the treatment of Newton [13] for both sections.

The discussion in the next two sections is supposed to be very general. In fact, even though the method described deals with the non-relativistic region, "the formalism set up is such that, provided there exists a consistent relativistic quantum field theory, the transition to the relativistic domain is relatively simple" [14]. However, we find that even in such a simple model as our Rearrangement Model, these anticipations are *not* fulfilled. This formalism leads to wrong and contradictory results, as will be discussed in section 13.

We wish to solve the Schrödinger equation

$$i\frac{\partial}{\partial t}\Psi\left(t\right) = H\Psi\left(t\right). \tag{2.1}$$

We split $H$ into a free Hamiltonian and an interaction Hamiltonian,

$$H = H_0 + H'. \tag{2.2}$$

We assume that this split can be carried out: we shall consider the case of rearrangement collisions later. We define four Green's functions

$$\left(i\frac{\partial}{\partial t} - H_0\right) G^{\pm}\left(t\right) = \mathbf{1}\delta\left(t\right), \tag{2.3$a$}$$

$$\left(i\frac{\partial}{\partial t} - H\right) \mathcal{G}^{\pm}\left(t\right) = \mathbf{1}\delta\left(t\right), \tag{2.3$b$}$$

with the initial conditions

$$G^{+}(t) = \mathcal{G}^{+}(t) = 0, \quad t < 0, \tag{2.4a}$$

$$G^{-}(t) = \mathcal{G}^{-}(t) = 0, \quad t > 0. \tag{2.4b}$$

$G^+$ and $\mathcal{G}^+$ are therefore the advanced Green's functions, and $G^-$ and $\mathcal{G}^-$ are the retarded Green's functions.

These may be solved formally yielding

$$G^{+}(t) = -ie^{-iH_0 t}\theta(t), \tag{2.5a}$$

$$G^{-}(t) = ie^{-iH_0 t}\theta(-t), \tag{2.5b}$$

$$\mathcal{G}^{+}(t) = -ie^{-iHt}\theta(t), \tag{2.5c}$$

$$\mathcal{G}^{-}(t) = ie^{-iHt}\theta(-t). \tag{2.5d}$$

Let $\Psi_0(t)$ be a state vector satisfying the free Schrödinger equation. The operator $G^+$ can then be used to express the state vector $\Psi_0(t')$ for any time $t'$ later than $t$, in terms of its value at $t' = t$,

$$\Psi_0(t') = iG^{+}(t'-t)\,\Psi_0(t). \tag{2.6}$$

$\Psi_0(t')$ then satisfies the free Schrödinger equation for $t' > t$, and $\Psi_0(t') \to \Psi_0(t)$ when $t' \to t$.

Therefore,

$$\lim_{t\to 0^+} G^{+}(t) = \lim_{t\to 0^+} \mathcal{G}^{+}(t) = -i\mathbf{1}, \tag{2.7a}$$

$$\lim_{t\to 0^-} G^{-}(t) = \lim_{t\to 0^-} \mathcal{G}^{-}(t) = i\mathbf{1}. \tag{2.7b}$$

Similarly, for $t' > t$ we can write

$$\Psi(t') = i\mathcal{G}^{+}(t'-t)\,\Psi(t), \tag{2.8}$$

and for $t' < t$ we have

$$\Psi_0\left(t'\right) = -iG^-\left(t'-t\right)\Psi_0\left(t\right), \tag{2.9a}$$

$$\Psi\left(t'\right) = -i\mathcal{G}^-\left(t'-t\right)\Psi\left(t\right). \tag{2.9b}$$

We now wish to define "in" and "out" states. We start by defining

$$\Psi_0\left(t\right) \equiv iG^+\left(t-t'\right)\Psi\left(t'\right), \tag{2.10}$$

whose time development for $t > t'$ is governed by the free Hamiltonian, but which at time $t_0$ was equal to $\Psi(t_0)$. We now let the time, $t'$, approach $\pm\infty$. Then, for the case of $t \to +\infty$, we have the "out" state, and for the case of $t \to -\infty$, we have the "in" state. In terms of the "in" and "out" states, the equations for $\Psi(t)$ are

$$\Psi\left(t\right) = \Psi_{\text{in}}\left(t\right) + \int_{-\infty}^{+\infty} dt'\,\mathcal{G}^+\left(t-t'\right)H'\Psi_{\text{in}}\left(t'\right), \tag{2.11a}$$

$$= \Psi_{\text{out}}\left(t\right) + \int_{-\infty}^{+\infty} dt'\,\mathcal{G}^-\left(t-t'\right)H'\Psi_{\text{in}}\left(t'\right). \tag{2.11b}$$

Note that these are retarded and advanced Green's functions for the *whole system.* These are *not* the same functions as those that appear in a (time ordered) Dyson series which are, instead, time ordered *particle propagators.* Note also that for every state in the continuous spectrum of $H_0$, and only for such states, these formulae define a corresponding state in the spectrum of $H$.

If we insert Eq. (2.9b) in Eq. (2.11a), we find

$$\Psi\left(t\right) = \Omega^{(+)}\Psi_{\text{in}}\left(t\right), \tag{2.12}$$

where

$$\begin{aligned}\Omega^{(+)} &= \mathbf{1} - i\int_{-\infty}^{+\infty} dt'\,\mathcal{G}^+\left(t-t'\right)H'G^-\left(t'-t\right)\\ &= \mathbf{1} - i\int_{-\infty}^{+\infty} dt\,\mathcal{G}^+\left(-t\right)H'G^-\left(t\right)\end{aligned} \tag{2.13}$$

is called the wave operator or the Möller matrix. We can similarly define $\Omega^{(-)}$.

Because $H'$ is hermitian, Eq. (2.13) gives us the relation

$$\Psi_{\text{in}}(t) = \Omega^{(+)\dagger}\Psi(t), \tag{2.14a}$$

and similarly

$$\Psi_{\text{out}}(t) = \Omega^{(-)\dagger}\Psi(t). \tag{2.14b}$$

Then, Eq. (2.12) and Eqs. (2.14) give us the relations

$$\Psi_{\text{in}}(t) = \Omega^{(+)\dagger}\Omega^{(+)}\Psi_{\text{in}}(t), \tag{2.15a}$$

$$\Psi_{\text{out}}(t) = \Omega^{(-)\dagger}\Omega^{(-)}\Psi_{\text{out}}(t). \tag{2.15b}$$

We now consider the possibility that the free states $\Psi_{\text{in}}(t)$ and $\Psi_{\text{out}}(t)$ span the entire Hilbert space, *i.e.* they are complete. From this assumption, we conclude that $\Omega^{(+)}$ and $\Omega^{(-)}$ are isometric,

$$\Omega^{(+)\dagger}\Omega^{(+)} = \Omega^{(-)\dagger}\Omega^{(-)} = \mathbf{1}. \tag{2.16}$$

This does not, however, mean that the $\Omega$'s are unitary: we cannot conclude from Eq. (2.15a) that Eq. (2.16) holds with its factors reversed.

Furthermore, granted the assumption that the $\Psi_{\text{in}}$ and $\Psi_{\text{out}}$ each form complete sets, we conclude that

$$H\Omega^{(\pm)} = \Omega^{(\pm)}H_0. \tag{2.17}$$

When there are bound states in the spectrum of $H$, we proceed as follows. Let $\Psi_0(E,\alpha)$ be the eigenstates of the free Hamiltonian with eigenvalue $E$, and $\alpha$ be the set of variables necessary to remove any degeneracy. Then, the completeness of these states can be written as a resolution of the identity,

$$\mathbf{1} = \sum_{\alpha}\int_0^{\infty} dE\,\Psi_0(E,\alpha)\,{\Psi_0}^{\dagger}(E,\alpha). \tag{2.18}$$

We again emphasize that we do not know whether the states of $H_0$ are complete, a priori. We are simply proceeding under that assumption. We then insert this into the product $\Omega\Omega^\dagger$, to get

$$\begin{aligned}\Omega\Omega^\dagger &= \Omega \int_0^\infty dE \sum_\alpha \Psi_0\left(E,\alpha\right) \Psi_0{}^\dagger\left(E,\alpha\right)\Omega^\dagger \\ &= \int_0^\infty dE \sum_\alpha \Psi\left(E,\alpha\right)\Psi^\dagger\left(E,\alpha\right) \\ &= \mathbf{1} - \Lambda. \end{aligned} \tag{2.19}$$

$\Lambda$ is called the unitary deficiency of $\Omega$. From the completeness of the set of all states, bound and scattering, of $H$,

$$\Lambda = \sum_n \Psi_{\rm bd}^{(n)} \Psi_{\rm bd}^{(n)\dagger}. \tag{2.20}$$

Thus, $\Lambda$ projects onto the space spanned by the bound states of $H$. If $H$ has no bound states, then $\Omega^{(+)}$ and $\Omega^{(-)}$ are unitary. Both $H$ and $H_0$ are hermitian; therefore, the hermitian conjugate of Eq. (2.17) gives

$$H_0\Omega^\dagger = \Omega^\dagger H. \tag{2.21}$$

We now let both sides of Eq. (2.21) act on $\Psi(E,\alpha)$ to get

$$H_0\Omega^\dagger\Psi\left(E,\alpha\right) = E\Omega^\dagger\Psi\left(E,\alpha\right), \tag{2.22}$$

which shows that if $E$ is in the spectrum of $H$ but not in the spectrum of $H_0$ then

$$\Omega^\dagger\Psi\left(E,\alpha\right) = 0,$$

and so

$$\Omega^\dagger\Lambda = 0. \tag{2.23}$$

Thus, the range of the operators $\Omega^{(\pm)}$ is not the entire Hilbert space. Instead, these operators map the whole space onto the subspace spanned by the continuum eigenstates of $H$. We cannot reach the subspace spanned by the bound states of $H$, and therefore, cannot construct an inverse operator for the whole space. The closest that we can come is to use the operators $\Omega^{(\pm)\dagger}$ which are inverses of $\Omega^{(\pm)}$ on the subspace of states spanned by the scattering states of $H$, and which annihilate the subspace of bound states of $H$.

Assuming that the asymptotic states are complete, we construct the S-matrix in the following manner. We use Eqs. (2.15b) and (2.12) to write the "out" state in terms of the "in" state,

$$\Psi_{\text{out}}(t) = \Omega^{(-)}\Omega^{(+)}\Psi_{\text{in}}(t)\,, \tag{2.24}$$

which defines for us the S-matrix

$$S \equiv \Omega^{(-)\dagger}\Omega^{(+)}. \tag{2.25}$$

The S-matrix can be shown to be unitary and isometric. See Newton [13] for details; note, however, that $S$ is only unitary when $\Omega$ is unitary. (This point is not clearly stated in Newton, or in the literature.)

Some mathematical questions about convergences arise in the above. Conventionally, in the Schrödinger picture (the one in which we are currently working), one demands that (see Newton [13] for details)

$$\lim_{t\to-\infty}\left[\Psi(t) - \psi_{\text{in}}(t)\right] \Rightarrow 0, \tag{2.26$a$}$$

$$\lim_{t\to+\infty}\left[\Psi(t) - \psi_{\text{out}}(t)\right] \Rightarrow 0, \tag{2.26$b$}$$

$$\lim_{t\to-\infty}\psi_{\text{in}}(0) \Rightarrow \Psi(0) \equiv \Omega^{(+)}\psi_{\text{in}}(0)\,, \tag{2.26$c$}$$

where $\Rightarrow$ denotes the strong limit.

We will find, in our model, that if we construct the asymptotic "in" and "out" states correctly, these limits will be satisfied; however, the states will not be orthonormal or complete. On the other hand, if we make the usual assumptions of scattering theory, namely that the asymptotic states are orthonormal and complete, then these limits will not be satisfied.

## 3. The Multiple Channel Formalism

The above formalism is only adequate for simple single-channel cases. For more general scattering problems, such as rearrangement collisions, we must generalize the formalism. We shall again follow Newton [13].

We want to split up the Hamiltonian into two pieces: one piece, $H_a$, that is left when the two initial fragments are taken far apart, and the remaining piece, $H'_a$. We can then go through the same development of $\Psi(t)$ from $\Psi_{\rm in}(t)$ as above. However, there is a difficulty that occurs for the development for the distant future. If rearrangements or break-ups can occur, then it is possible that the "channel" Hamiltonian in the future is different than the "channel" Hamiltonian in the past.

The various possibilities for an $n$-particle system are handled by defining a partition of them into $k$ clusters, denoted by $a_k$. Given a partition $a$, we define $H_a$ by allowing all distances between clusters to independently tend to infinity. Therefore, $H_a$ will contain only interactions that are internal to clusters, but none between them. Then, $H'_a$ is defined by the requirement that $H = H_a + H'_a$, and therefore, for any two partitions $a$ and $b$, we have

$$H = H_a + H'_a = H_b + H'_b. \tag{3.1}$$

To each partition, there will correspond Green's functions given by

$$\left(i\frac{\partial}{\partial t} - H_a\right) G^{\pm}{}_a(t) = \mathbf{1}\delta(t), \tag{3.2}$$

with the same boundary conditions as Eqs. (2.4). If $H_a$, after removing the kinetic energy of the center of mass motion and of the centers of mass of its clusters, has at least one bound state, it is called an arrangement channel. When this condition on $H_a$ does not hold, the channel is not of interest as an initial or final scattering state. If $H_a$ has more than one bound state, then each of them defines a separate channel, and therefore, in each channel the clusters are in a specific bound state but moving freely relative to one another. The channel consisting of the entire $n$-cluster partition is the channel 0 because then $H_a = H_0$.

Now consider the space of each arrangement channel $a$, which we shall denote by $\mathcal{H}_a$. Then, if $a$ has $m$ fragments, each state in $\mathcal{H}_a$ will have $m$ groups of bound particles. This means that unless the channel $a$ is the entire $n$-fragment arrangement channel, $\mathcal{H}_a$ will not be the whole Hilbert space: the ionized eigenstates of $H_a$ will be missing. Furthermore, as each $\mathcal{H}_a$ is defined by different channel Hamiltonians, $H_a$, the $\mathcal{H}_a$'s are generally not orthogonal to each other. In fact, "the complete set of basis functions is not linearly independent and, of course, not orthonormal" [15].

It will be convenient to define the orthogonal projections $P_a$ onto the channel spaces, $\mathcal{H}_a$. In other words, we define

$$P_a^2 = P_a, \quad P_a^{\dagger} = P_a, \quad P_a\mathcal{H}_a = \mathcal{H}_a, \tag{3.3}$$

with the null space of $P_a$ defined as the space spanned by the ionized eigenstates of $H_a$. $P_a$ projects states from the full Hilbert space, $\mathcal{H}$, to the channel spaces, $\mathcal{H}_a$. Obviously, for the $n$-cluster arrangement channel, we have $P_0 = \mathbf{1}$.

We now wish to define “in” and “out” states. We first define an $a$ state, which is a state that develops according to $H_a$ but is in $\mathcal{H}_a$,

$$\left(i\frac{\partial}{\partial t} - H_a\right)\Psi_a(\alpha, t) = 0, \tag{3.4}$$

where the label $\alpha$ contains all the other information including the arrangement channel (even though including the arrangement channel in $\alpha$ is redundant for $\Psi_a(\alpha, t)$, it is convenient for other purposes).

We then define $\Psi^{(+)}(\alpha, t)$ as a state in $\mathcal{H}$ that develops according to $H$,

$$\left(i\frac{\partial}{\partial t} - H\right)\Psi^{(+)}(\alpha, t) = 0, \tag{3.5}$$

and for which there exists an $a$-state such that

$$\lim_{t\to-\infty}\left(\Psi_a(\alpha, t), \Psi^{(+)}(\alpha, t)\right) = 1. \tag{3.6}$$

Therefore, the state $\Psi_a(\alpha, t)$ is the “in” state $\Psi_{\text{in}}(\alpha, t)$ in relation to the state $\Psi^{(+)}(\alpha, t)$. Eq. (3.6) demands that the probability of finding the system in state $\Psi_a(\alpha, t)$ in the remote past approach 1, and therefore, it is equivalent to

$$\Psi_a^{(+)} \underset{t\to-\infty}{\rightrightarrows} \Psi_a(\alpha, t), \tag{3.7$a$}$$

or

$$\int dt\, iG_a^+(t - t')\,\Psi^{(+)}(\alpha, t') \underset{t\to-\infty}{\rightrightarrows} \Psi_{\text{in}}(\alpha, t), \tag{3.7$b$}$$

with the double arrow denoting the strong limit.

Similarly,

$$\int dt'\,(-i)\,G_a^-(t - t')\,\Psi^{(-)}(\alpha, t') \underset{t\to-\infty}{\rightrightarrows} \Psi_{\text{out}}(\alpha, t). \tag{3.8}$$

Exactly analogous to Eqs. (2.11), we can now write

$$\Psi^{(+)}(\alpha,t) = \Psi_{\rm in}(\alpha,t) + \int_{-\infty}^{+\infty} dt'\,\mathcal{G}^{+}\left(t-t'\right) H_a' \Psi_{\rm in}\left(\alpha,t'\right), \tag{3.9$a$}$$

$$\Psi^{(-)}(\alpha,t) = \Psi_{\rm out}(\alpha,t) + \int_{-\infty}^{+\infty} dt'\,\mathcal{G}^{-}\left(t-t'\right) H_a' \Psi_{\rm out}\left(\alpha,t'\right). \tag{3.9$b$}$$

We can now define the Möller matrices, and the S-matrix. The Möller matrices are defined by

$$\Psi^{\pm}(\alpha,t) = \Omega_a^{(\pm)} \Psi_a(\alpha,t), \tag{3.10}$$

with only those states $\Psi_a$ admitted which are in $\mathcal{H}_a$. On the orthogonal complement (*i.e.* the ionized eigenstates of $H_a$) $\Omega^{(\pm)}$ is defined to be zero,

$$\Omega^{(\pm)} P_a = \Omega^{(\pm)}.$$

Then, on the space $\mathcal{H}_a$, using Eq. (3.9a), we find

$$\begin{aligned} \Omega_a^{(+)} &= P_a + K_a^{(+)}, \\ K_a^{(+)} &= -i \int_{-\infty}^{\infty} dt\,\mathcal{G}^{+}(-t)\, H_a' G^{-}{}_a(t)\, P_a \\ &= -i \int_{-\infty}^{0} dt\, e^{iHt} H_a' e^{-iH_a t} P_a, \end{aligned}$$

and therefore,

$$\Omega_a^{(+)} = \lim_{t\to-\infty} e^{iHt} e^{-iH_a t} P_a. \tag{3.11}$$

We can similarly find, on $\mathcal{H}_a$, that

$$\Omega_a^{(-)} = \lim_{t\to\infty} e^{iHt} e^{-iH_a t} P_a. \tag{3.12}$$

The range of $\Omega_a^{(+)}$ is the space of all full states that develop from arrangement channel $a$, and the range of $\Omega_a^{(-)}$ is the space of all full states that develop into arrangement channel $a$. Let us call these ranges $\mathcal{R}_a^{(+)}$ and $\mathcal{R}_a^{(-)}$, and their respective orthogonal projections $Q_a^{(+)}$ and $Q_a^{(-)}$. The Möller matrices, $\Omega^{(\pm)}$, map $\mathcal{H}_a$ onto $\mathcal{R}_a^{\pm}$, and from Eqs. (3.10) we find that on $\mathcal{R}_a^{(+)}$ and $\mathcal{R}_a^{(-)}$, respectively,

$$\begin{aligned} \Psi_a(\alpha, t) &= \Psi_{\text{in}}(\alpha, t) = \Omega^{(+)\dagger}\Psi^{(+)}(\alpha, t) \\ &= \Psi_{\text{out}}(\alpha, t) = \Omega^{(-)\dagger}\Psi^{(-)}(\alpha, t). \end{aligned} \tag{3.13}$$

Therefore, because the $\Psi_a(\alpha, t)$ span the space $\mathcal{H}_a$, we find that the Möller matrices, $\Omega_a^{(\pm)}$, are partially isometric from the space $\mathcal{H}_a$, *i.e.*

$$\Omega_a^{(\pm)\dagger}\Omega_a^{(\pm)} = P_a. \tag{3.14}$$

Similarly, the $\Omega_a^{(\pm)\dagger}$ are partially isometric from the ranges $\mathcal{R}_a^{(\pm)}$ of the $\Omega^{(\pm)}$, *i.e.*

$$\Omega_a^{(\pm)}\Omega_a^{(\pm)\dagger} = Q_a^{(\pm)}, \tag{3.15}$$

which defines the $Q_a^{(\pm)}$. The full states developing from or into any arrangement channel are orthogonal to each other as can be seen by direct evaluation of the inner products of asymptotic states. "If the two arrangement channels are different, then there must be at least one particle for which the "overlap" of the two states was negligible in the remote past because it belonged to a different fragment. Hence that inner product must vanish for all times" [16].

A major point of difference with our results from the Rearrangement Model is the statement, "note that the same argument shows that the inner product

$$(\Psi_b(\beta, t), \Psi_a(\alpha, t)) \tag{3.16}$$

approaches zero as $t \to \pm\infty$ (for $a \neq b$). But since $H_a \neq H_b$, it is not independent of $t$ and hence it does not generally vanish for *finite* times" (emphasis added) [16]. In the Rearrangement Model, this is untrue: we show in section 7 that our states are all orthogonal to each other.

From the Schrödinger equation, one can write

$$H\Omega_a^{(\pm)} = \Omega_a^{(\pm)} H_a, \tag{3.17}$$

which means that $\Omega$ intertwines $H$ and $H_a$. This again is a major difference with the Rearrangement Model, because we show in section 7 that $\Omega$ intertwines $H$ and $H_C$, where $H_C$ is the comparison Hamiltonian, which has the same spectrum as $H$; here, $H_a$ does not have the same spectrum as $H$.

Our channel definitions could also include the single cluster arrangement channel, which is the channel of all the $n$-particle bound states of $H$. If we define $\Lambda$ to be the orthogonal projection onto that subspace, then for all $a$ we have

$$Q_a^{(\pm)} \Lambda = 0. \tag{3.18}$$

Now, every non-bound state must be decomposable into states that arise from, or go into, one of the other arrangements. Therefore, we assume

$$\Lambda + \sum_a Q_a^{(\pm)} = \mathbf{1}, \tag{3.19}$$

which is known as asymptotic completeness.

Using Eqs. (3.12), (3.17), and (3.19), we may then define a unitary S-matrix,

$$\Psi^{(+)}(\alpha, t) \quad \underset{t \to -\infty}{\Rightarrow} \quad \Psi_{\text{out}}(t) = \sum_b S_{ba} \Psi_a(\alpha, t), \tag{3.20}$$

where

$$S_{ba} = \Omega_b^{(-)\dagger} \Omega_a^{(+)}. \tag{3.21}$$

The mathematical questions of convergence are the same here as for the single channel case, as in Eqs. (2.26).

## 4. The Rearrangement Model

To keep contact with earlier work, we shall use a combination of the notations of [7] and [17], as far as possible. We consider a quantum field theory with five distinct fields, $B$, $C$, $D$, $\theta$, and $\phi$, and the corresponding particles (no anti-particles).

The non-zero commutators are:

$$\begin{aligned} &\left[B, B^{\dagger}\right] = \left[D, D^{\dagger}\right] = \left[C, C^{\dagger}\right] = 1, \\ &\left[\theta(\omega), \theta^{\dagger}(\omega')\right] = \delta\left(\omega' - \omega\right), \quad \left[\phi(\nu), \phi^{\dagger}(\nu')\right] = \delta\left(\nu' - \nu\right). \end{aligned} \tag{4.1}$$

Note that $\theta$ and $\phi$ are labelled by continuum parameters, $0 < \omega, \nu < \infty$, while $B$, $C$, and $D$, are treated as single modes ("infinitely heavy") [18]. We choose to use the energy as our variable, rather than momentum, because this makes the model much simpler, and more physically transparent. We want a total Hamiltonian for the system which allows the transitions

$$B \leftrightarrow C\phi,$$

and

$$D \leftrightarrow C\theta.$$

Therefore, we choose our Hamiltonian to be:

$$H = H_0 + V, \tag{4.2}$$

where

$$H_0 = m_B B^\dagger B + m_D D^\dagger D + \int d\omega\, \omega \theta^\dagger(\omega)\theta(\omega) + \int d\nu\, \nu \phi^\dagger(\nu)\phi(\nu), \tag{4.3}$$

and

$$\begin{aligned} V = &\int d\omega\, f(\omega)\, \theta(\omega) C D^\dagger + \int d\omega\, f^*(\omega)\, \theta^\dagger(\omega) C^\dagger D \\ &+ \int d\nu\, g(\nu)\, \phi(\nu) C B^\dagger + \int d\nu\, g^*(\nu)\, \phi^\dagger(\nu) C^\dagger B. \end{aligned} \tag{4.4}$$

This Hamiltonian has three constants of motion apart from itself. They are:

$$C_1 = B^\dagger B + C^\dagger C + D^\dagger D, \tag{4.5$a$}$$

$$C_2 = B^\dagger B + \int d\nu\, \phi^\dagger(\nu)\phi(\nu), \tag{4.5$b$}$$

$$C_3 = D^\dagger D + \int d\omega\, \theta^\dagger(\omega)\theta(\omega). \tag{4.5$c$}$$

Therefore, no transitions can occur between sectors labelled by different values of these quantum numbers. Let us start by enumerating the stable sectors. The first such sector is the vacuum and has $C_1 = C_2 = C_3 = 0$. The next three are: $C_1 = 1, C_2 = 0, C_3 = 0$; $C_1 = 0, C_2 = 1, C_3 = 0$; and $C_1 = 0, C_2 = 0, C_3 = 1$. These correspond to the states $C$, $\phi$, and $\theta$, respectively. Finally, there is the sector with $C_1 = 0, C_2 = 1, C_3 = 1$; it corresponds to a state $\theta\phi$.

The three lowest non-trivial sectors are:

$$C_1 = 1, C_2 = 1, C_3 = 0, \tag{4.6$a$}$$

$$C_1 = 1, C_2 = 0, C_3 = 1, \tag{4.6$b$}$$

$$C_1 = 1, C_2 = 1, C_3 = 1. \tag{4.6$c$}$$

These correspond to $B \leftrightarrow C\phi$, $D \leftrightarrow C\theta$, and $B\theta \leftrightarrow C\theta\phi \leftrightarrow D\phi$, respectively. The last of these is the sector in which rearrangement collisions can take place, and as mentioned before, we shall call this the Rearrangement Sector.

Our strategy for solving the model in the Rearrangement Sector will be to first construct the solutions of the two lowest non-trivial sectors, (4.6a) and (4.6b) (which are exactly analogous to the Lee Model), and then use these solutions to express the Rearrangement Sector equations.

## 5. Solving the Model

We start by constructing the solutions for the $B \leftrightarrow C\phi$ and $D \leftrightarrow C\theta$ sectors. These are exactly the same as the Lee model, so the solutions are simple. We shall denote non-interacting ("bare") states by single bras and kets ($\langle\ ,\ \rangle$), and interacting states ("dressed" or "physical") by double bras and kets ($\langle\langle\ ,\ \rangle\rangle$).

The equations we need to solve for the continuum solutions are

$$H|\lambda\rangle\rangle = \lambda|\lambda\rangle\rangle \tag{5.1$a$}$$

in the $B \leftrightarrow C\phi$ sector, and

$$H|\mu\rangle\rangle = \mu|\mu\rangle\rangle \tag{5.1$b$}$$

in the $D \leftrightarrow C\theta$ sector.

We define

$$\alpha\,(z) \equiv z - m_D - \int_0^\infty d\omega\, \frac{|f\,(\omega)\,|^2}{z-\omega}, \tag{5.2}$$

$$\beta\,(z) \equiv z - m_B - \int_0^\infty d\nu\, \frac{|g\,(\nu)\,|^2}{z-\nu}, \tag{5.3}$$

$$\rho_{\lambda,B}(\nu) \equiv \langle C\phi(\nu)|\lambda\rangle\rangle, \tag{5.4}$$

$$\rho_{\mu,D}(\omega) \equiv \langle C\theta(\omega)|\mu\rangle\rangle, \tag{5.5}$$

$$\sigma_{\lambda,B} \equiv \langle B|\lambda\rangle\rangle, \tag{5.6}$$

$$\sigma_{\mu,D} \equiv \langle D|\mu\rangle\rangle. \tag{5.7}$$

For shorthand, we will write $\alpha(\lambda)$ for $\alpha(\lambda+i\epsilon)$ and $\alpha^*(\lambda)$ for $\alpha(\lambda-i\epsilon)$, and similarly for $\beta(\lambda)$. In terms of these, the solutions are:

$$\rho_{\lambda,B}(\nu) = \delta\left(\lambda-\nu\right) + \frac{g^*\left(\nu\right)\sigma_{\lambda,B}}{\lambda-\nu+i\epsilon}, \tag{5.8$a$}$$

$$\rho_{\mu,D}(\omega) = \delta\left(\mu-\omega\right) + \frac{f^*\left(\omega\right)\sigma_{\mu,D}}{\mu-\omega+i\epsilon}, \tag{5.8$b$}$$

$$\sigma_{\lambda,B} = \frac{g\left(\lambda\right)}{\beta\left(\lambda\right)}, \tag{5.8$c$}$$

$$\sigma_{\mu,D} = \frac{f\left(\mu\right)}{\alpha\left(\mu\right)}. \tag{5.8$d$}$$

If $\alpha(z)$ develops zeros then we have additional discrete states. Similarly, if $\beta(z)$ develops zeros then again we have additional discrete states. For our purposes, we shall always assume that both $\alpha(z)$ and $\beta(z)$ have exactly one zero each, which are denoted by $M_D$ and $M_B$, respectively. There is no loss of generality if we use this assumption because the extension to more than one zero is trivial. The equations for the discrete states are:

$$H|M_B\rangle\rangle = M_B|M_B\rangle\rangle \tag{5.9$a$}$$

in the $B \leftrightarrow C\phi$ sector, and

$$H|M_D\rangle\rangle = M_D|M_D\rangle\rangle \tag{5.9$b$}$$

in the $D \leftrightarrow C\theta$ sector. We define

$$\rho_B(\nu) \equiv \langle C\phi(\nu)|M_B\rangle\rangle, \tag{5.10}$$

$$\rho_D(\omega) \equiv \langle C\theta(\omega)|M_D\rangle\rangle, \tag{5.11}$$

$$\sqrt{Z_B} \equiv \langle B|M_B\rangle\rangle, \tag{5.12}$$

$$\sqrt{Z_D} \equiv \langle D|M_D\rangle\rangle. \tag{5.13}$$

In terms of these, the normalized solutions are:

$$\rho_B(\nu) = \sqrt{Z_B}\frac{g^*(\nu)}{M_B - \nu}, \tag{5.14a}$$

$$\rho_D(\omega) = \sqrt{Z_D}\frac{f^*(\omega)}{M_D - \omega}, \tag{5.14b}$$

$$Z_B = \left[1 + \int d\nu \, \frac{|g(\nu)|^2}{(M_B - \nu)^2}\right]^{-1}, \tag{5.14c}$$

$$Z_D = \left[1 + \int d\omega \, \frac{|f(\omega)|^2}{(M_D - \omega)^2}\right]^{-1}, \tag{5.14d}$$

where the last two are obtained by imposition of the orthonormality condition. Note that these solutions, Eqs. (5.8) and (5.14), form a complete orthonormal set.

Now, we use these solutions to construct the solutions in the Rearrangement Sector. In this sector we will have four sorts of solutions. The first will correspond to the "physical" $|C\theta(\omega)\phi(\nu)\rangle\rangle$ sector, the second to the "physical" $|D\phi(\nu)\rangle\rangle$, the third to the "physical" $|B\theta(\omega)\rangle\rangle$, and the last to one or more dynamically generated bound states, which we shall denote by $|M_A\rangle\rangle$. Which solution is obtained will depend on whether we put the delta functions (which represent the plane wave parts of our solutions) in our solutions. If we put none, we get the discrete states.

We wish to solve the eigenvalue equation

$$H|E\rangle\rangle = E|E\rangle\rangle. \tag{5.15}$$

We expand Eq. (5.15) in terms of $\theta^\dagger(\omega)|\lambda\rangle\rangle$, $\theta^\dagger(\omega)|M_B\rangle\rangle$, $\phi^\dagger(\nu)|\mu\rangle\rangle$, and $\phi^\dagger(\nu)|M_D\rangle\rangle$. By acting on these states with the Hamiltonian, and using Eqs. (5.1) and (5.9) we get

$$(E - \lambda - \omega)\langle\langle\lambda|\theta(\omega)|E\rangle\rangle = f^*(\omega)\,\langle\langle\lambda|C^\dagger D|E\rangle\rangle, \tag{5.16a}$$

$$(E - M_B - \omega)\langle\langle M_B|\theta(\omega)|E\rangle\rangle = f^*(\omega)\,\langle\langle M_B|C^\dagger D|E\rangle\rangle, \tag{5.16b}$$

$$(E - \mu - \nu)\langle\langle\mu|\phi(\nu)|E\rangle\rangle = g^*(\nu)\,\langle\langle\mu|C^\dagger B|E\rangle\rangle, \tag{5.16c}$$

$$(E - M_D - \nu)\langle\langle M_D|\phi(\nu)|E\rangle\rangle = g^*(\nu)\,\langle\langle M_D|C^\dagger B|E\rangle\rangle. \tag{5.16$d$}$$

We need to evaluate the unknown matrix elements on the right hand side of Eqs. (5.16). We solve for these elements by inserting $H$ in them, commuting it on one side, and letting it act on $|E\rangle\rangle$ on the other. For example, we can solve for $\langle\langle\lambda|C^\dagger D|E\rangle\rangle$ in the following manner:

$$\begin{aligned}
&\langle\langle\lambda|C^\dagger DH|E\rangle\rangle = E\langle\langle\lambda|C^\dagger D|E\rangle\rangle \\
&\quad\Rightarrow \langle\langle\lambda|\Big\{HC^\dagger D + \Big[C^\dagger D, H\Big]\Big\}|E\rangle\rangle = E\langle\langle\lambda|C^\dagger D|E\rangle\rangle.
\end{aligned} \tag{5.17}$$

We now let $H$ in the first term of Eq. (5.17) act on $\langle\langle\lambda|$, and evaluate the commutator in the second term. We proceed similarly for the other three equations and, when all the dust settles, get

$$\begin{aligned}
&(E - \lambda - m_D)\,\langle\langle\lambda|CD^\dagger|E\rangle\rangle \\
&\quad = \int d\omega\, f(\omega)\,\langle\langle\lambda|\theta(\omega)|E\rangle\rangle \\
&\quad - \sigma_{\lambda,B}{}^*\Big[\int d\omega\, f(\omega)\Big\{\int d\lambda'\sigma_{\lambda',B}\langle\langle\lambda'|\theta(\omega)|E\rangle\rangle + \sqrt{Z_B}\langle\langle M_B|\theta(\omega)|E\rangle\rangle\Big\} \\
&\qquad + \int d\nu\, g(\nu)\Big\{\int d\mu'\sigma_{\mu',D}\langle\langle\mu'|\phi(\nu)|E\rangle\rangle + \sqrt{Z_D}\langle\langle M_D|\phi(\nu)|E\rangle\rangle\Big\}\Big],
\end{aligned} \tag{5.18$a$}$$

$$\begin{aligned}
&(E - M_B - m_D)\,\langle\langle M_B|CD^\dagger|E\rangle\rangle \\
&\quad = \int d\omega\, f(\omega)\,\langle\langle M_B|\theta(\omega)|E\rangle\rangle \\
&\quad - \sqrt{Z_B}\Big[\int d\omega\, f(\omega)\Big\{\int d\lambda'\sigma_{\lambda',B}\langle\langle\lambda'|\theta(\omega)|E\rangle\rangle + \sqrt{Z_B}\langle\langle M_B|\theta(\omega)|E\rangle\rangle\Big\} \\
&\qquad + \int d\nu\, g(\nu)\Big\{\int d\mu'\sigma_{\mu',D}\langle\langle\mu'|\phi(\nu)|E\rangle\rangle + \sqrt{Z_D}\langle\langle M_D|\phi(\nu)|E\rangle\rangle\Big\}\Big],
\end{aligned} \tag{5.18$b$}$$

$$\begin{aligned}
&(E - \mu - m_B)\,\langle\langle\mu|CB^\dagger|E\rangle\rangle \\
&\quad = \int d\nu\, g(\nu)\,\langle\langle\mu|\phi(\nu)|E\rangle\rangle \\
&\quad - \sigma_{\mu,D}{}^*\Big[\int d\omega\, f(\omega)\Big\{\int d\lambda'\sigma_{\lambda',B}\langle\langle\lambda'|\theta(\omega)|E\rangle\rangle + \sqrt{Z_B}\langle\langle M_B|\theta(\omega)|E\rangle\rangle\Big\}
\end{aligned}$$

$$+\int d\nu\, g(\nu)\left\{\int d\mu' \sigma_{\mu',D}\langle\langle\mu'|\phi(\nu)|E\rangle\rangle + \sqrt{Z_D}\langle\langle M_D|\phi(\nu)|E\rangle\rangle\right\}\Bigg], \quad (5.18c)$$

$$(E - M_D - m_B)\langle\langle M_D|CB^\dagger|E\rangle\rangle$$
$$= \int d\nu\, g(\nu)\langle\langle M_D|\phi(\nu)|E\rangle\rangle$$
$$- \sqrt{Z_D}\Bigg[\int d\omega\, f(\omega)\left\{\int d\lambda' \sigma_{\lambda',B}\langle\langle\lambda'|\theta(\omega)|E\rangle\rangle + \sqrt{Z_B}\langle\langle M_B|\theta(\omega)|E\rangle\rangle\right\}$$
$$+\int d\nu\, g(\nu)\left\{\int d\mu' \sigma_{\mu',D}\langle\langle\mu'|\phi(\nu)|E\rangle\rangle + \sqrt{Z_D}\langle\langle M_D|\phi(\nu)|E\rangle\rangle\right\}\Bigg]. \quad (5.18d)$$

We first solve for the "physical" $|C\theta(\omega)\phi(\nu)\rangle\rangle$ states. We start by inverting Eqs. (5.16) and putting in the requisite delta functions. Note that we may put the product of two delta functions because this is an infinitely degenerate (double) continuum, which cannot be specified by just $E$; rather, we have to label the state with the variables $E$ and $n$, with the $n$ variable representing the division of energy between the $\theta$ and $\phi$ particles. We then substitute these into the first term of each of Eqs. (5.18), and solve for the unknown matrix elements. Having found them, we put them into Eqs. (5.16) to find our solutions. Defining

$$b^C(E,n,\lambda,\omega) \equiv \langle\langle\lambda|\theta(\omega)|E,n\rangle\rangle, \quad (5.19a)$$

$$b_F^C(E,n,M_B,\omega) \equiv \langle\langle M_B|\theta(\omega)|E,n\rangle\rangle, \quad (5.19b)$$

$$d^C(E,n,\mu,\nu) \equiv \langle\langle\mu|\phi(\nu)|E,n\rangle\rangle, \quad (5.19c)$$

$$d_F^C(E,n,M_D,\nu) \equiv \langle\langle M_D|\phi(\nu)|E,n\rangle\rangle, \quad (5.19d)$$

we get

$$b^C(E,n,\lambda,\omega) = \delta(E-\lambda-n)\rho_{n,D}(\omega) - \frac{f^*(\omega)}{(E-\lambda-\omega+i\epsilon)}\frac{{\sigma_{\lambda,B}}^*}{\alpha(E-\lambda)}K_C(E,n), \quad (5.20a)$$

$$b_F^C(E,n,M_B,\omega) = -\frac{f^*(\omega)}{(E-M_B-\omega+i\epsilon)}\frac{\sqrt{Z_B}}{\alpha(E-M_B)}K_C(E,n), \quad (5.20b)$$

$$d^C(E,n,\mu,\nu) = \delta(\mu-n)\rho_{E-n,B}(\nu) - \frac{g^*(\nu)}{(E-\mu-\nu+i\epsilon)}\frac{{\sigma_{\mu,D}}^*}{\beta(E-\mu)}K_C(E,n), \quad (5.20c)$$

$$d_F^C(E,n,M_D,\nu) = -\frac{g^*(\nu)}{(E-M_D-\nu+i\epsilon)}\frac{\sqrt{Z_D}}{\beta(E-M_D)}K_C(E,n), \quad (5.20d)$$

where

$$K_C\left(E,n\right)=\frac{g\left(E-n\right)f\left(n\right)}{\beta\left(E-n\right)\alpha\left(n\right)}\frac{1}{\gamma\left(E\right)},\tag{5.21}$$

$$\gamma\left(E\right)=\frac{Z_D}{\beta\left(E-M_D\right)}+\int d\mu\,\frac{\left|f\left(\mu\right)\right|^2}{\left|\alpha\left(\mu\right)\right|^2}\frac{1}{\beta\left(E-\mu\right)}\tag{5.22a}$$

$$=\frac{Z_B}{\alpha\left(E-M_B\right)}+\int d\lambda\,\frac{\left|g\left(\lambda\right)\right|^2}{\left|\beta\left(\lambda\right)\right|^2}\frac{1}{\alpha\left(E-\lambda\right)}.\tag{5.22b}$$

The last equality follows from Eqs. (A.10).

We now solve for the "physical" $|D\phi(\nu)\rangle\rangle$ sector. We must again start by inverting Eqs. (5.16), but this time need to put just the one requisite delta function in Eq. (5.16d). We put these equations in Eqs. (5.18), and solve for the unknown matrix elements putting in another delta function in Eq. (5.18a) when inverting because now we must account for the zero of $\alpha(E-\lambda)$ at $M_D$. We then put these results in Eqs. (5.16). Defining

$$b^D(E,\lambda,\omega)\equiv\langle\langle\lambda|\theta(\omega)|E\rangle\rangle,\tag{5.23a}$$

$$b_F^D(E,M_B,\omega)\equiv\langle\langle M_B|\theta(\omega)|E\rangle\rangle,\tag{5.23b}$$

$$d^D(E,\mu,\nu)\equiv\langle\langle\mu|\phi(\nu)|E\rangle\rangle,\tag{5.23c}$$

$$d_F^D(E,M_D,\nu)\equiv\langle\langle M_D|\phi(\nu)|E\rangle\rangle,\tag{5.23d}$$

we get

$$b^D(E,\lambda,\omega)=\rho_D\left(\omega\right)\delta(E-\lambda-M_D)-\frac{f^*\left(\omega\right)}{\left(E-\lambda-\omega+i\epsilon\right)}\frac{{\sigma_{\lambda,B}}^*}{\alpha\left(E-\lambda\right)}K_D\left(E\right),\tag{5.24a}$$

$$b_F^D(E,M_B,\omega)=-\frac{f^*\left(\omega\right)}{\left(E-M_B-\omega+i\epsilon\right)}\frac{\sqrt{Z_B}}{\alpha\left(E-M_B\right)}K_D\left(E\right),\tag{5.24b}$$

$$d^D(E,\mu,\nu)=-\frac{g^*\left(\nu\right)}{\left(E-\mu-\nu+i\epsilon\right)}\frac{{\sigma_{\mu,D}}^*}{\beta\left(E-\mu\right)}K_D\left(E\right),\tag{5.24c}$$

$$d_F^D(E,M_D,\nu)=\rho_{E-M_D,B}\left(\nu\right)-\frac{g^*\left(\nu\right)}{\left(E-M_D-\nu+i\epsilon\right)}\frac{\sqrt{Z_D}}{\beta\left(E-M_D\right)}K_D\left(E\right),\tag{5.24d}$$

where

$$K_D\left(E\right)=\frac{\sqrt{Z_D}}{\gamma\left(E\right)}\frac{g\left(E-M_D\right)}{\beta\left(E-M_D\right)},\tag{5.25}$$

and $\gamma(E)$ is the same as that defined in Eqs. (5.22).

In exactly the same way, we can find the solutions for the "physical" $|B\theta(\omega)\rangle\rangle$ sector. They are:

$$b^B(E,\lambda,\omega) = -\frac{f^*(\omega)}{(E-\lambda-\omega+i\epsilon)}\frac{{\sigma_{\lambda,B}}^*}{\alpha(E-\lambda)}K_B(E), \tag{5.26a}$$

$$b^B_F(E,M_B,\omega) = \rho_{E-M_B,D}(\omega) - \frac{f^*(\omega)}{(E-M_B-\omega+i\epsilon)}\frac{\sqrt{Z_B}}{\alpha(E-M_B)}K_B(E), \tag{5.26b}$$

$$d^B(E,\mu,\nu) = \rho_B(\nu)\,\delta(E-\mu-M_B) - \frac{g^*(\nu)}{(E-\mu-\nu+i\epsilon)}\frac{{\sigma_{\mu,D}}^*}{\beta(E-\mu)}K_B(E), \tag{5.26c}$$

$$d^B_F(E,M_D,\nu) = -\frac{g^*(\nu)}{(E-M_D-\nu+i\epsilon)}\frac{\sqrt{Z_D}}{\beta(E-M_D)}K_B(E), \tag{5.26d}$$

where

$$K_B(E) = \frac{\sqrt{Z_B}}{\gamma(E)}\frac{f(E-M_B)}{\alpha(E-M_B)}, \tag{5.27}$$

and $\gamma(E)$ is the same as that defined in Eq. (5.22).

Finally, we wish to solve for any dynamically generated discrete states. In this case, we put no delta functions anywhere. When we follow the procedure of putting Eqs. (5.16) in Eqs. (5.18) and solving for the unknown matrix elements, we find that the only way to satisfy all the equations is if $\gamma(E)$ has zeros. Denoting these zeros by $M_A$, we find the discrete state solutions:

$$b^A(M_A,\lambda,\omega) = -\frac{f^*(\omega)}{(M_A-\lambda-\omega)}\frac{{\sigma_{\lambda,B}}^*}{\alpha(M_A-\lambda)}K_A(M_A), \tag{5.28a}$$

$$b^A_F(M_A,M_B,\omega) = -\frac{f^*(\omega)}{(M_A-M_B-\omega+i\epsilon)}\frac{\sqrt{Z_B}}{\alpha(M_A-M_B)}K_A(M_A), \tag{5.28b}$$

$$d^A(M_A,\mu,\nu) = -\frac{g^*(\nu)}{(M_A-\mu-\nu)}\frac{{\sigma_{\mu,D}}^*}{\beta(M_A-\mu)}K_A(M_A), \tag{5.28c}$$

$$d^A_F(M_A,M_D,\nu) = -\frac{g^*(\nu)}{(M_A-M_D-\nu+i\epsilon)}\frac{\sqrt{Z_D}}{\beta(M_A-M_D)}K_A(M_A), \tag{5.28d}$$

where $K_A(M_A)$ is now an arbitrary normalization factor which is fixed, when demonstrating completeness, to be $\sqrt{\frac{d\gamma(E)}{dE}|_{E=M_A}}$ (see the discussion after Eq. (7.18)). For our purposes, without loss of generality, we assume that there is only one zero of $\gamma(E)$, denoted by $M_A$, and thus only one dynamically generated discrete state. The extension to more than one discrete state is trivial.

In each of Eqs. (5.20), (5.24), (5.26), and (5.28) the superscript refers to the sector that the solution is in, and the subscript $F$ refers to solutions expanded in the the discrete state part of the Lee Model sectors. Furthermore, we have anticipated future developments by fixing the arbitrary constants accompanying the delta functions in Eqs. (5.20), (5.24), and (5.26). We do this by demanding that Eq. (6.3a) and Eq. (6.3b), or their equivalents for the other two sectors, give the same result, and that the solutions be orthonormal.

## 6. Verification of the Solutions

We now proceed to verify that Eqs. (5.20), (5.24), (5.26), and (5.28) are each solutions to our problem. To do this, we first transform our solutions into the bare state basis; *i.e.* in terms of the non-interacting states $|C\theta(\omega)\phi(\nu)\rangle$, $|B\theta(\omega)\rangle$, and $|D\phi(\nu)\rangle$, using the completeness of the lower sector solutions. With the expansion coefficients in the "physical" $|C\theta(\omega)\phi(\nu)\rangle\rangle$ sector defined in the following manner (with the coefficients for the other sectors defined similarly)

$$|E,n\rangle\rangle \equiv C^C(E,n,\omega,\nu)|C\theta(\omega)\phi(\nu)\rangle + B^C(E,n,\omega)|B\theta(\omega)\rangle + D^C(E,n,\nu)|D\phi(\nu)\rangle, \tag{6.1}$$

where

$$\Psi^C(E,n,\omega,\nu) \equiv \begin{pmatrix} C^C(E,n,\omega,\nu) \\ D^C(E,n,\nu) \\ B^C(E,n,\omega) \end{pmatrix}, \tag{6.2}$$

and

$$C^C(E,n,\omega,\nu) \equiv \langle C\theta(\omega)\phi(\nu)|E,n\rangle\rangle$$

$$B^C(E,n,\omega) \equiv \langle B\theta(\omega)|E,n\rangle\rangle$$

$$D^C(E,n,\nu) \equiv \langle D\phi(\nu)|E,n\rangle\rangle.$$

we have

$$C^C(E,n,\omega,\nu) = \int d\lambda\, \rho_{\lambda,B}(\nu) b^C(E,n,\lambda,\omega) + \rho_B(\nu) b_F^C(E,n,M_B,\omega), \qquad (6.3a)$$

$$= \int d\mu\, \rho_{\mu,D}(\omega) d^C(E,n,\mu,\nu) + \rho_D(\omega) d_F^C(E,n,M_D,\nu), \qquad (6.3b)$$

$$D^C(E,n,\nu) = \int d\mu\, \sigma_{\mu,D} d^C(E,n,\mu,\nu) + \sqrt{Z_D} d_F^C(E,n,M_D,\nu), \qquad (6.3c)$$

$$B^C(E,n,\omega) = \int d\lambda\, \sigma_{\lambda,B} b^C(E,n,\lambda,\omega) + \sqrt{Z_B} b_F^C(E,n,M_B,\omega), \qquad (6.3d)$$

with similar equations for the other three sectors (for example, for the "physical" $|D\phi(\nu)\rangle\rangle$ sector, we would replace $C^C(E,n,\omega,\nu)$ by $C^D(E,\omega,\nu)$, $b^C(E,n,\lambda,\omega)$ by $b^D(E,\lambda,\omega)$, etc.). A good check that we have solved our equations correctly is to verify that Eqs. (6.3a) and (6.3b) give the same result. This is indeed completely trivial if we use Eq. (A.14).

For the "physical" $|C\theta(\omega)\phi(\nu)\rangle\rangle$ sector in the bare basis, we get:

$$C^C(E,n,\omega,\nu) = \rho_{n,D}(\omega)\rho_{E-n,B}(\nu) - K_C(E,n) \frac{f^*(\omega)\, g^*(\nu)}{(E-\omega-\nu+i\epsilon)}$$
$$\left\{ \int d\lambda \frac{|\sigma_{\lambda,B}|^2}{(E-\lambda-\omega+i\epsilon)\alpha(E-\lambda)} + \int d\mu \frac{|\sigma_{\mu,D}|^2}{(E-\mu-\nu+i\epsilon)\beta(E-\mu)} \right.$$
$$\left. + \frac{Z_B}{\alpha(E-M_B)(E-M_B-\omega+i\epsilon)} + \frac{Z_D}{\beta(E-M_D)(E-M_D-\nu+i\epsilon)} \right\}, \qquad (6.4a)$$

$$D^C(E,n,\nu) = \frac{f(n)}{\alpha(n)}\rho_{n,D}(\omega) - K_C(E,n)\, g^*(\nu)$$
$$\left\{ \int d\mu \frac{|\sigma_{\mu,D}|^2}{(E-\mu-\nu+i\epsilon)\beta(E-\mu)} + \frac{Z_D}{\beta(E-M_D)(E-M_D-\nu+i\epsilon)} \right\}, \qquad (6.4b)$$

$$B^C(E,n,\omega) = \frac{g(E-n)}{\beta(E-n)}\rho_{E-n,B}(\nu) - K_C(E,n)\, f^*(\omega)$$
$$\left\{ \int d\lambda \frac{|\sigma_{\lambda,B}|^2}{(E-\lambda-\omega+i\epsilon)\alpha(E-\lambda)} + \frac{Z_B}{\alpha(E-M_B)(E-M_B-\omega+i\epsilon)} \right\}. \qquad (6.4c)$$

For the "physical" $|D\phi(\nu)\rangle\rangle$ sector in the bare basis, we get:

$$C^D(E,\omega,\nu) = \rho_D(\omega)\,\rho_{E-M_D,B}(\nu) - K_D(E)\,\frac{f^*(\omega)\,g^*(\nu)}{(E-\omega-\nu+i\epsilon)}$$
$$\left\{\int d\lambda\,\frac{|\sigma_{\lambda,B}|^2}{(E-\lambda-\omega+i\epsilon)\alpha(E-\lambda)} + \int d\mu\,\frac{|\sigma_{\mu,D}|^2}{(E-\mu-\nu+i\epsilon)\beta(E-\mu)}\right.$$
$$\left.+\frac{Z_B}{\alpha(E-M_B)(E-M_B-\omega+i\epsilon)} + \frac{Z_D}{\beta(E-M_D)(E-M_D-\nu+i\epsilon)}\right\}, \quad (6.5a)$$

$$D^D(E,\nu) = \sqrt{Z_D}\rho_{E-M_D,B}(\nu) - K_D(E)\,g^*(\nu)$$
$$\left\{\int d\mu\,\frac{|\sigma_{\mu,D}|^2}{(E-\mu-\nu+i\epsilon)\beta(E-\mu)} + \frac{Z_D}{\beta(E-M_D)(E-M_D-\nu+i\epsilon)}\right\}, \quad (6.5b)$$

$$B^D(E,\omega) = \rho_D(\omega)\,\sigma_{E-M_D,B} - K_D(E)\,f^*(\omega)$$
$$\left\{\int d\lambda\,\frac{|\sigma_{\lambda,B}|^2}{(E-\lambda-\omega+i\epsilon)\alpha(E-\lambda)} + \frac{Z_B}{\alpha(E-M_B)(E-M_B-\omega+i\epsilon)}\right\}. \quad (6.5c)$$

For the "physical" $|B\theta(\omega)\rangle\rangle$ sector in the bare basis, we get:

$$C^B(E,\omega,\nu) = \rho_B(\nu)\,\rho_{E-M_B,D}(\omega) - K_B(E)\,\frac{f^*(\omega)\,g^*(\nu)}{(E-\omega-\nu+i\epsilon)}$$
$$\left\{\int d\lambda\,\frac{|\sigma_{\lambda,B}|^2}{(E-\lambda-\omega+i\epsilon)\alpha(E-\lambda)} + \int d\mu\,\frac{|\sigma_{\mu,D}|^2}{(E-\mu-\nu+i\epsilon)\beta(E-\mu)}\right.$$
$$\left.+\frac{Z_B}{\alpha(E-M_B)(E-M_B-\omega+i\epsilon)} + \frac{Z_D}{\beta(E-M_D)(E-M_D-\nu+i\epsilon)}\right\}, \quad (6.6a)$$

$$D^B(E,\nu) = \rho_B(\nu)\,\sigma_{E-M_B,D} - K_B(E)\,g^*(\nu)$$
$$\left\{\int d\mu\,\frac{|\sigma_{\mu,D}|^2}{(E-\mu-\nu+i\epsilon)\beta(E-\mu)} + \frac{Z_D}{\beta(E-M_D)(E-M_D-\nu+i\epsilon)}\right\}, \quad (6.6b)$$

$$B^B(E,\omega) = \sqrt{Z_B}\rho_{E-M_B,D}(\omega) - K_B(E)\,f^*(\omega)$$
$$\left\{\int d\lambda\,\frac{|\sigma_{\lambda,B}|^2}{(E-\lambda-\omega+i\epsilon)\alpha(E-\lambda)} + \frac{Z_B}{\alpha(E-M_B)(E-M_B-\omega+i\epsilon)}\right\}. \quad (6.6c)$$

Finally, for the discrete states, we get:

$$C^A(M_A,\omega,\nu) = -K_A(M_A)\,\frac{f^*(\omega)\,g^*(\nu)}{M_A-\omega-\nu}$$
$$\left\{\int d\lambda\,\frac{|\sigma_{\lambda,B}|^2}{(M_A-\lambda-\omega)\alpha(M_A-\lambda)} + \int d\mu\,\frac{|\sigma_{\mu,D}|^2}{(M_A-\mu-\nu)\beta(M_A-\mu)}\right.$$
$$+\frac{Z_B}{\alpha(M_A-M_B)(M_A-M_B-\omega+i\epsilon)}$$

$$+\frac{Z_D}{\beta\left(M_A-M_D\right)\left(M_A-M_D-\nu+i\epsilon\right)}\Bigg\}, \tag{6.7a}$$

$$D^A\left(M_A,\nu\right)=-K_A\left(M_A\right)g^*\left(\nu\right)$$

$$\left\{\int d\mu\,\frac{|\sigma_{\mu,D}|^2}{(M_A-\mu-\nu)\beta\left(M_A-\mu\right)}+\frac{Z_D}{\beta\left(M_A-M_D\right)\left(M_A-M_D-\nu+i\epsilon\right)}\right\}, \tag{6.7b}$$

$$B^A\left(M_A,\omega\right)=-K_A\left(M_A\right)f^*\left(\omega\right)$$

$$\left\{\int d\lambda\,\frac{|\sigma_{\lambda,B}|^2}{(M_A-\lambda-\omega)\alpha\left(M_A-\lambda\right)}+\frac{Z_B}{\alpha\left(M_A-M_B\right)\left(M_A-M_B-\omega+i\epsilon\right)}\right\}. \tag{6.7c}$$

We now verify that Eqs. (6.4), (6.5), (6.6), and (6.7) are each solutions of our model. To do this, we explicitly write down the analogues of Eqs. (5.16) in the bare basis, plug in each set of solutions in turn, and verify that the equations are satisfied. A straightforward analysis shows that the following equations must be satisfied in the bare basis (we have written them for the "physical" $|C\theta(\omega)\phi(\nu)\rangle\rangle$ sector, *i.e.* with the variable $n$—for the other sectors the variable $n$ is, of course, missing):

$$(E-\omega-\nu)C\left(E,n,\omega,\nu\right)=g^*\left(\nu\right)B\left(E,n,\omega\right)+f^*\left(\omega\right)D\left(E,n,\nu\right), \tag{6.8a}$$

$$\left(E-m_B-\omega\right)B\left(E,n,\omega\right)=\int d\nu\,g\left(\nu\right)C\left(E,n,\omega,\nu\right), \tag{6.8b}$$

$$\left(E-m_D-\nu\right)D\left(E,n,\nu\right)=\int d\omega\,f\left(\omega\right)C\left(E,n,\omega,\nu\right). \tag{6.8c}$$

Putting each of Eqs. (6.4), (6.5), (6.6), and (6.7) in turn into Eqs. (6.8), or their equivalents for the other sectors, and using Eq. (A.14), we find that each of these sets of solutions satisfies the equations. Incidentally, a glance at Eqs. (6.8) immediately shows why we could not have solved the problem directly rather than the somewhat convoluted method we went through: the integral equations are not separable, and are quite intractable.

## 7. Orthonormality and Completeness

We now proceed to verify orthonormality and completeness of the solutions Eqs. (6.4), (6.5), (6.6), and (6.7). We start by verifying orthonormality for the diagonal components beginning with the scalar product $(\Psi^{C^\dagger}(E', n', \omega, \nu), \Psi^{C}(E, n, \omega, \nu))$, which is given by

$$\begin{aligned}\int d\omega\, d\nu\, \Psi^{C^\dagger}(E', n', \omega, \nu)\Psi^{C}(E, n, \omega, \nu) \\ = \int d\omega\, d\nu\, C^{C^*}(E', n', \omega, \nu)C^{C}(E, n, \omega, \nu) \\ + \int d\omega\, B^{C^*}(E', n', \omega)B^{C}(E, n, \omega) \\ + \int d\nu\, D^{C^*}(E', n', \nu)D^{C}(E, n, \nu).\end{aligned} \tag{7.1}$$

We now use Eqs. (6.3) to write this as

$$\begin{aligned}\int d\omega\, d\nu\, \Psi^{C^\dagger}(E', n', \omega, \nu)\Psi^{C}(E, n, \omega, \nu) \\ = \int d\omega\, d\nu \left[\int d\lambda'\, \rho^*_{\lambda',B}(\nu) b^{C^*}(E', n', \lambda', \omega) + \rho^*_B(\nu) b_F^{C^*}(E', n', M_B, \omega)\right] \\ \left[\int d\lambda\, \rho_{\lambda,B}(\nu) b^{C}(E, n, \lambda, \omega) + \rho_B(\nu) b_F^{C}(E, n, M_B, \omega)\right] \\ + \int d\omega \left[\int d\lambda'\, \sigma^*_{\lambda',B} b^{C^*}(E', n', \lambda', \omega) + \sqrt{Z_B} b_F^{C^*}(E', n', M_B, \omega)\right] \\ \left[\int d\lambda\, \sigma_{\lambda,B} b^{C}(E, n, \lambda, \omega) + \sqrt{Z_B} b_F^{C}(E, n, M_B, \omega)\right] \\ + \int d\nu\, D^{C^*}(E', n', \nu)D^{C}(E, n, \nu).\end{aligned} \tag{7.2}$$

We then do the integrals over $\lambda$ and $\lambda'$ to find

$$\begin{aligned}\int d\omega\, d\nu\, \Psi^{C^\dagger}(E', n', \omega, \nu)\Psi^{C}(E, n, \omega, \nu) \\ = \int d\lambda\, d\omega\, b^{C^*}(E', n', \lambda, \omega) b^{C}(E, n, \lambda, \omega) \\ + \int d\omega\, b_F^{C^*}(E', n', M_B, \omega) b_F^{C}(E, n, M_B, \omega) \\ + \int d\nu\, D^{C^*}(E', n', \nu)D^{C}(E, n, \nu).\end{aligned} \tag{7.3}$$

Defining

$$L_1\left(E',n'\right) \equiv \frac{f^*\left(n'\right) g^*\left(E'-n'\right)}{\alpha^*\left(n'\right)\beta^*\left(E'-n'\right)},$$
$$L_2\left(E,n\right) \equiv \frac{f\left(n\right) g\left(E-n\right)}{\alpha\left(n\right)\beta\left(E-n\right)}, \tag{7.4}$$

we find that the sum of the first two integrals is

$$\begin{aligned}
&\delta\left(E-E'\right)\delta\left(n-n'\right) - \delta\left(E'-n'-E+n\right)\frac{f^*\left(n'\right) f\left(n\right)}{\alpha^*\left(n'\right)\alpha\left(n\right)} \\
&+ L_1\left(E',n'\right) L_2\left(E,n\right)\left\{\frac{1}{\gamma^*\left(E'\right)\alpha^*\left(E'-E+n\right)} + \frac{1}{\gamma\left(E\right)\alpha\left(E-E'+n'\right)}\right\} \\
&+ \frac{L_1\left(E',n'\right) L_2\left(E,n\right)}{\gamma^*\left(E'\right)\gamma\left(E\right)}\left\{\frac{-Z_B}{\alpha^*\left(E'-M_B\right)\alpha\left(E-M_B\right)} - \int d\lambda\,\frac{\left|\sigma_{\lambda,B}\right|^2}{\alpha^*\left(E'-\lambda\right)\alpha\left(E-\lambda\right)}\right\},
\end{aligned} \tag{7.5}$$

while the third integral gives

$$\begin{aligned}
&\delta\left(E'-n'-E+n\right)\frac{f^*\left(n'\right) f\left(n\right)}{\alpha^*\left(n'\right)\alpha\left(n\right)} \\
&- L_1\left(E',n'\right) L_2\left(E,n\right)\left\{\frac{1}{\gamma^*\left(E'\right)\alpha^*\left(E'-E+n\right)} + \frac{1}{\gamma\left(E\right)\alpha\left(E-E'+n'\right)}\right\} \\
&+ \frac{L_1\left(E',n'\right) L_2\left(E,n\right)}{\gamma^*\left(E'\right)\gamma\left(E\right)} \\
&\qquad\left\{\frac{Z_D}{\beta\left(E-M_D\right)\alpha^*\left(E'-E+M_D\right)} + \frac{Z_D}{\beta^*\left(E'-M_D\right)\alpha\left(E-E'+M_D\right)}\right. \\
&\qquad\left. + \int d\mu'\,\frac{\left|\sigma_{\mu',D}\right|^2}{\beta^*\left(E'-\mu'\right)\alpha\left(E-E'+\mu'\right)} + \int d\mu\frac{\left|\sigma_{\mu,D}\right|^2}{\beta\left(E-\mu\right)\alpha^*\left(E'-E-\mu\right)}\right\}.
\end{aligned} \tag{7.6}$$

Adding Eqs. (7.5) and (7.6) together, and doing the integrals by combining them into a single contour integral (which evaluates simply to its residues), we find that the only term left is $\delta(E'-E)\delta(n'-n)$, which is just as required.

We can similarly show that

$$\begin{aligned}
&\int d\omega\, d\nu\; \Psi^{D\dagger}(E',\omega,\nu)\Psi^D(E,\omega,\nu) \\
&\qquad = \int d\lambda\, d\omega\; b^{D^*}(E',\lambda,\omega) b^D(E,\lambda,\omega)
\end{aligned}$$

$$+\int d\omega\, b_F^{D*}(E',M_B,\omega)b_F^D(E,M_B,\omega)$$

$$+\int d\nu\, D^{D*}(E',\nu)D^D(E,\nu)$$

$$=\delta\left(E'-E\right), \tag{7.7}$$

and

$$\int d\omega\, d\nu\, \Psi^{B\dagger}(E',\omega,\nu)\Psi^B(E,\omega,\nu)$$

$$=\int d\lambda\, d\omega\, b^{B*}(E',\lambda,\omega)b^B(E,\lambda,\omega)$$

$$+\int d\omega\, b_F^{B*}(E',M_B,\omega)b_F^B(E,M_B,\omega)$$

$$+\int d\nu\, D^{B*}(E',\nu)D^B(E,\nu)$$

$$=\delta\left(E'-E\right). \tag{7.8}$$

Finally,

$$\Psi^{A\dagger}(M_A,\omega,\nu)\Psi^A(M_A,\omega,\nu)=1. \tag{7.9}$$

Now we take up the off-diagonal elements. For

$$\int d\omega\, d\nu\, \Psi^{C\dagger}(E',n',\omega,\nu)\Psi^D(E,\omega,\nu)$$

$$=\int d\lambda\, d\omega\, b^{C*}(E',n',\lambda,\omega)b^D(E,\lambda,\omega)$$

$$+\int d\omega\, b_F^{C*}(E',n',M_B,\omega)b_F^D(E,M_B,\omega)$$

$$+\int d\nu\, D^{C*}(E',n',\nu)D^D(E,\nu), \tag{7.10}$$

we find that the third integral exactly cancels the sum of the first two, giving us 0. We can similarly show that

$$\int d\omega\, d\nu\, \Psi^{C\dagger}(E',n',\omega,\nu)\Psi^B(E,\omega,\nu)=0, \tag{7.11}$$

$$\int d\omega\, d\nu\, \Psi^{C\dagger}(E',n',\omega,\nu)\Psi^A(M_A,\omega,\nu)=0, \tag{7.12}$$

$$\int d\omega\, d\nu\, \Psi^{D^\dagger}(E',\omega,\nu)\Psi^B(E,\omega,\nu) = 0, \tag{7.13}$$

$$\int d\omega\, d\nu\, \Psi^{D^\dagger}(E',\omega,\nu)\Psi^A(M_A,\omega,\nu) = 0, \tag{7.14}$$

$$\int d\omega\, d\nu\, \Psi^{B^\dagger}(E',\omega,\nu)\Psi^A(M_A,\omega,\nu) = 0. \tag{7.15}$$

Therefore, the set of solutions we found, Eqs. (6.4), (6.5), (6.6), and (6.7) are orthonormal.

We now consider completeness. We wish to show that

$$\begin{aligned}
&\int dE\, dn\, \Psi^C(E,n,\omega,\nu)\Psi^{C^\dagger}(E,n,\omega',\nu') + \int dE\, \Psi^D(E,\omega,\nu)\Psi^{D^\dagger}(E,\omega',\nu') \\
&+ \int dE\, \Psi^B(E,\omega,\nu)\Psi^{B^\dagger}(E,\omega',\nu') + \Psi^A(M_A,\omega,\nu)\Psi^{A^\dagger}(M_A,\omega',\nu') \\
&= \begin{pmatrix} \delta(\nu-\nu')\,\delta(\omega-\omega') & 0 & 0 \\ 0 & \delta(\nu-\nu') & 0 \\ 0 & 0 & \delta(\omega-\omega') \end{pmatrix}.
\end{aligned} \tag{7.16}$$

Let us start with the diagonal elements. The $(1,1)$ element of the matrix is

$$\begin{aligned}
&\int dE\, dn\, C^C(E,n,\omega,\nu)C^{C^*}(E,n,\omega',\nu') + \int dE\, C^D(E,\omega,\nu)C^{D^*}(E,\omega',\nu') \\
&+ \int dE\, C^B(E,\omega,\nu)C^{B^*}(E,\omega',\nu') + C^A(M_A,\omega,\nu)C^{A^*}(M_A,\omega',\nu').
\end{aligned} \tag{7.17}$$

These integrals are most easily done in the following manner. The first term can be rewritten, using Eqs. (6.3), as

$$\begin{aligned}
&\int dEdn\, C^C(E,n,\omega,\nu)C^{C^*}(E,n,\omega',\nu') \\
&\quad \left\{ \int d\lambda'\, \rho^*_{\lambda',B}(\nu') b^{C^*}(E,n,\lambda',\omega') + \rho^*_B(\nu') b_F^{C^*}(E,n,M_B,\omega') \right\} \\
&\quad = \int dE\, dn \left\{ \int d\lambda\, \rho_{\lambda,B}(\nu) b^C(E,n,\lambda,\omega) + \rho_B(\nu) b_F^C(E,n,M_B,\omega) \right\}.
\end{aligned} \tag{7.18}$$

One then rewrites subsequent terms in Eq. (7.17) in a similar fashion as Eq. (7.18). Since the integrals are exceedingly tedious, we describe how they are done, and leave it to the interested reader to verify our results. The integrals over $n$ are done with the help of Eq. (A.15). Then, the integrals over $E$ are done by converting them into contour integrals. When all the contour integrals are combined it is found that they add together into one large contour integral (plus the non-contributing circle at infinity), which evaluates simply to its residues. These residues exactly cancel the other pieces in the expression, leaving over one or more delta functions for the diagonal terms, and nothing for the off-diagonal ones. For convenience, the branch cuts and poles of the function $\frac{1}{\gamma(z)}$, where $z$ is a complex integration variable in the contour integral, are shown in Figure 3.

One finds that the $(1,1)$ term is $\delta(\omega-\omega')\delta(\nu-\nu')$. In doing this, one has to fix $K_A(M_A)=\sqrt{\frac{d\gamma(E)}{dE}|_{E=M_A}}$, which fixes the unknown normalization constant in Eqs. (5.28). One can similarly show that the $(2,2)$ and the $(3,3)$ terms are $\delta(\nu-\nu')$ and $\delta(\omega-\omega')$, respectively.

For the off-diagonal terms, one proceeds similarly and finds that they are all zero. Thus, our set of solutions, namely, Eqs. (6.4), (6.5), (6.6), and (6.7) is a complete orthonormal set of solutions of our model in this sector.

## 8. The Möller Matrix and the Comparison Hamiltonian

The matrix (with continuous eigenvalues) of the eigenfunctions, including any discrete solutions, gives us the generalized Möller Matrix by virtue of the results already demonstrated on orthonormality and completeness [7]. It is given by

$$\Omega\left(E,n,\omega,\nu\right)=\left(\Psi^C(E,n,\omega,\nu),\Psi^D(E,\omega,\nu),\Psi^B(E,\omega,\nu),\Psi^A(M_A,\omega,\nu)\right) \tag{8.1}$$

with components

$$\begin{pmatrix} C^C(E,n,\omega,\nu) & C^D(E,\omega,\nu) & C^B(E,\omega,\nu) & C^A(M_A,\omega,\nu) \\ D^C(E,n,\nu) & D^D(E,\nu) & D^B(E,\nu) & D^A(M_A,\nu) \\ B^C(E,n,\omega) & B^D(E,\omega) & B^B(E,\omega) & B^A(M_A,\omega) \end{pmatrix}. \tag{8.2}$$

It has the properties of being unitary

$$\Omega\Omega^\dagger = \mathbf{1}, \tag{8.3$a$}$$

$$\Omega^\dagger\Omega = \mathbf{1}, \tag{8.3$b$}$$

and of diagonalizing the full Hamiltonian, $H$,

$$H\Omega = \Omega H_C, \tag{8.4$a$}$$

$$\Omega^\dagger H\Omega = H_C, \tag{8.4$b$}$$

where $H_C$ is called the comparison Hamiltonian. It can be calculated in the following manner. First, we use the eigenvalue equations to write

$$\begin{aligned} &H\Omega\,(E,n,\omega,\nu) \\ &\quad = \Big(E\Psi^C(E,n,\omega,\nu), E\Psi^D(E,\omega,\nu), E\Psi^B(E,\omega,\nu), M_A\Psi^A(M_A,\omega,\nu)\Big)\,, \end{aligned} \tag{8.5}$$

and then act on Eq. (8.5) with $\Omega^\dagger$ from the left, and make use of the orthonormality relations to get

$$\begin{aligned} \Omega^\dagger H\Omega &= \begin{pmatrix} E\delta\,(E-E')\,\delta\,(n-n') & 0 & 0 & 0 \\ 0 & E\delta\,(E-E') & 0 & 0 \\ 0 & 0 & E\delta\,(E-E') & 0 \\ 0 & 0 & 0 & M_A \end{pmatrix} \\ &= H_C. \end{aligned} \tag{8.6}$$

To compare $H_C$ with the free Hamiltonian, $H_0$, we rewrite $H_C$, putting $E = n + \tau$ for the $(1,1)$ element, $E = M_D + \tau$ for the $(2,2)$ element, and $E = M_B + \tau$ for the $(3,3)$ element, and similarly for $E'$. Thus, $H_C$ becomes

$$\begin{pmatrix} (n+\tau)\,\delta(\tau-\tau')\,\delta(n-n') & 0 & 0 & 0 \\ 0 & (M_D+\tau)\,\delta(\tau-\tau') & 0 & 0 \\ 0 & 0 & (M_B+\tau)\,\delta(\tau-\tau') & 0 \\ 0 & 0 & 0 & M_A \end{pmatrix}. \tag{8.7}$$

The free Hamiltonian, $H_0$, is

$$\begin{pmatrix} (\omega+\nu)\,\delta(\omega-\omega')\,\delta(\nu-\nu') & 0 & 0 \\ 0 & (m_D+\nu)\,\delta(\nu-\nu') & 0 \\ 0 & 0 & (m_B+\omega)\,\delta(\omega-\omega') \end{pmatrix}. \tag{8.8}$$

Comparing $H_C$ and $H_0$, we see that we can identify $H_C$ with $H_0$ if we include *both* mass and wave-function renormalization terms in the interaction, and ignore the discrete $M_A$ state in $H_C$. The mass renormalization means that we must add a quantity $\Delta$ to $H_0$, where $\Delta$ is

$$\Delta = \begin{pmatrix} 0 & 0 & 0 \\ 0 & (M_D-m_D)\,\delta(\nu-\nu') & 0 \\ 0 & 0 & (M_B-m_B)\,\delta(\omega-\omega') \end{pmatrix}. \tag{8.9}$$

The structure of our solutions, Eqs. (6.4), (6.5), and (6.6) immediately tells us that we must have a wave function (and consequent coupling constant) renormalization.

Thus, the fields $B$, $C$, $D$, $\theta$, and $\phi$ have the wave function renormalizations

$$B \to \sqrt{\beta'}B = \frac{1}{\sqrt{Z_B}}B, \tag{8.10$a$}$$

$$D \to \sqrt{\alpha'}D = \frac{1}{\sqrt{Z_D}}D, \tag{8.10$b$}$$

$$C \to C, \tag{8.10$c$}$$

$$\theta \to \theta, \tag{8.10$d$}$$

$$\phi \to \phi. \tag{8.10$e$}$$

Because there are no proper vertex corrections, the coupling constant renormalizations reflect the wave function renormalizations [7]

$$f(\omega) \to \sqrt{Z_D}f(\omega), \tag{8.11}$$

$$g(\nu) \to \sqrt{Z_B}g(\nu). \tag{8.12}$$

Furthermore, as there are no divergences in this problem, the coupling constant and wave function renormalizations are inessential, and the mass renormalization making $H_0 + \Delta$ identifiable with $H_C$ is essential only in this sector. These renormalizations are sufficient for higher sectors as well. The only change in the higher sectors is due to the mass renormalizations which alter the continuum thresholds from $m_D$ and $m_B$ to $M_D$ and $M_B$, respectively, but leave everything else unaffected.

Notice that while $H_C$ and $H_0$ have the same structure (as long as $\alpha$ and $\beta$ both have zeros, and $\gamma$ does not), they have different spectra. Only the double continuum $0 < n < E < \infty$ is coextensive; the $D\phi$ and $B\theta$ continua are renormalized downwards from $m_D$ to $M_D$ and from $m_B$ to $M_B$, respectively. Notice also that, contrary to conventional wisdom [1,3,12], the Möller matrix intertwines the full Hamiltonian, $H$, with $H_C$, *not* with $H_0$. However, $H_C$ and $H$ do have the same spectrum.

In addition, if we take the unitary transformation of $H_C$ in reverse, we can convert the comparison Hamiltonian to the full Hamiltonian

$$\Omega H_C \Omega^\dagger = H, \tag{8.13}$$

and just as in the Cascade Model of [7], we find that *the notion of an interaction is basis dependent.*

## 9. The S Matrix

We have obtained one set of solutions to our problem, namely, Eqs. (6.4), (6.5), (6.6), and (6.7). We can, of course, obtain another set in which the singular operators of the form $(E-\omega-\nu+i\epsilon)^{-1}$ (which give the "in" states) in Eqs. (6.4), (6.5), and (6.6) are changed to $(E-\omega-\nu-i\epsilon)^{-1}$ (which give the "out" states), while Eqs. (6.7) remain unchanged. Let us denote these solutions, and quantities associated with them, with a prime. This new set also furnishes a Möller matrix,

$$\Omega' = \left(\Psi^{C'}(E,n,\omega,\nu), \Psi^{D'}(E,\omega,\nu), \Psi^{B'}(E,\omega,\nu), \Psi^{A}(M_A,\omega,\nu)\right), \tag{9.1}$$

which satisfies the same properties as the original Möller matrix, that of unitarity

$$\Omega'\Omega'^{\dagger} = \mathbf{1}, \tag{9.2$a$}$$

$$\Omega'^{\dagger}\Omega' = \mathbf{1}, \tag{9.2$b$}$$

and of diagonalizing $H$

$$H\Omega' = \Omega' H_C, \tag{9.3$a$}$$

$$\Omega'^{\dagger} H\Omega' = H_C. \tag{9.3$b$}$$

The set of states, Eqs. (6.4), (6.5), (6.6), and (6.7) are such that

$$\lim_{t\to-\infty} e^{iH_C t} e^{-iHt}\Psi^C(E,n,\omega,\nu) = \begin{pmatrix} \delta(n-\omega)\,\delta(E-\omega-\nu) \\ 0 \\ 0 \end{pmatrix}, \tag{9.4$a$}$$

$$\lim_{t\to-\infty} e^{iH_C t} e^{-iHt}\Psi^D(E,\omega,\nu) = \begin{pmatrix} 0 \\ \sqrt{Z_D}\,\delta(E-M_D-\nu) \\ 0 \end{pmatrix}, \tag{9.4$b$}$$

$$\lim_{t\to-\infty} e^{iH_C t} e^{-iHt}\Psi^B(E,\omega,\nu) = \begin{pmatrix} 0 \\ 0 \\ \sqrt{Z_B}\,\delta(E-M_B-\omega) \end{pmatrix}, \tag{9.4$c$}$$

$$\lim_{t\to-\infty} e^{iH_C t} e^{-iHt}\Psi^A(M_A,\omega,\nu) = \Psi^A(M_A,\omega,\nu), \tag{9.4$d$}$$

of which the first three are the plane wave ideal eigenstates of the comparison Hamiltonian. However, notice that there is the need for a *wave function renormalization* in $\Psi^D(E,\omega,\nu)$ and $\Psi^B(E,\omega,\nu)$, and that the *threshold is renormalized* in these two cases (*i.e.* $m_B \to M_B$ and $m_D \to M_D$). Clearly, these states are the "in" states in our problem. This is again analogous to the Cascade Model of [7].

For $t \to +\infty$ for these "in" states we have

$$\lim_{t\to+\infty} e^{iH_C t} e^{-iHt} \Psi^C(E,n,\omega,\nu)$$
$$= \begin{pmatrix} \delta(E-\omega-\nu)\left[\delta(n-\omega)\frac{\alpha^*(n)\beta^*(E-n)}{\alpha(n)\beta(E-n)} + \frac{2\pi i}{\gamma(E)}\frac{f(n)g(E-n)}{\alpha(n)\beta(E-n)}\frac{f^*(\omega)g^*(\nu)}{\alpha(\omega)\beta(\nu)}\right] \\ \frac{2\pi i}{\gamma(E)}\frac{f(n)g(E-n)}{\alpha(n)\beta(E-n)} Z_D \delta(E-M_D-\nu)\frac{g^*(\nu)}{\beta(\nu)} \\ \frac{2\pi i}{\gamma(E)}\frac{f(n)g(E-n)}{\alpha(n)\beta(E-n)} Z_B \delta(E-M_B-\omega)\frac{f^*(\omega)}{\alpha(\omega)} \end{pmatrix}, \quad (9.5a)$$

$$\lim_{t\to+\infty} e^{iH_C t} e^{-iHt} \Psi^D(E,\omega,\nu)$$
$$= \begin{pmatrix} 2\pi i\delta(E-\omega-\nu)\frac{\sqrt{Z_D}}{\gamma(E)}\frac{g(E-M_D)}{\beta(E-M_D)}\frac{f^*(\omega)g^*(E-\omega)}{\alpha(\omega)\beta(E-\omega)} \\ \sqrt{Z_D}\delta(E-M_D-\nu)\left[\frac{\beta^*(\nu)}{\beta(\nu)} + \frac{2\pi i}{\gamma(E)} Z_D \frac{|g(\nu)|^2}{\beta(\nu)\beta(\nu)}\right] \\ \sqrt{Z_B}\delta(E-M_B-\omega)\frac{2\pi i}{\gamma(E)} Z_B \frac{|f(\omega)|^2}{\alpha(\omega)\alpha(\omega)} \end{pmatrix}, \quad (9.5b)$$

$$\lim_{t\to+\infty} e^{iH_C t} e^{-iHt} \Psi^B(E,\omega,\nu)$$
$$= \begin{pmatrix} 2\pi i\delta(E-\omega-\nu)\frac{\sqrt{Z_B}}{\gamma(E)}\frac{f(E-M_B)}{\alpha(E-M_B)}\frac{f^*(\omega)g^*(E-\omega)}{\alpha(\omega)\beta(E-\omega)} \\ \sqrt{Z_D}\delta(E-M_D-\nu)\frac{2\pi i}{\gamma(E)} Z_D \frac{|g(\nu)|^2}{\beta(\nu)\beta(\nu)} \\ \sqrt{Z_B}\delta(E-M_B-\omega)\left[\frac{\alpha^*(\omega)}{\alpha(\omega)} + \frac{2\pi i}{\gamma(E)} Z_B \frac{|f(\omega)|^2}{\alpha(\omega)\alpha(\omega)}\right] \end{pmatrix}, \quad (9.5c)$$

$$\lim_{t\to+\infty} e^{iH_C t} e^{-iHt} \Psi^A(M_A,\omega,\nu) = \Psi^A(M_A,\omega,\nu). \quad (9.5d)$$

(The limits in Eqs. (9.4) and Eqs. (9.5) are understood for multiplication by smooth functions of $\omega$ or $\nu$ or both, as the case may be).

The "out" states behave in an analogous but opposite fashion to the "in" states. They behave simply for $t \to +\infty$, but have a complicated structure as $t \to -\infty$. Furthermore, the "in" states at $t \to -\infty$ and the "out" states at $t \to +\infty$ are identical. Therefore, we can define an S-matrix, and can compute it in one of several ways. For example, we can compute it using

$$\Psi_{\text{scattered}} = \lim_{t\to\infty} \left(\Psi(t) - \Psi(-t)\right), \tag{9.6}$$

or we can take the scalar product of the "in" and "out" states

$$\left(\Psi', \Psi\right) = S. \tag{9.7}$$

Both methods, of course, give the same answer.

The method of the inner products is cleaner and more aesthetically satisfying so we shall follow it for the calculation. The results are easily checked by doing the calculation by the other methods.

Schematically, the S-matrix looks like (the "+" subscript means an "in" state and the "−" subscript means an "out" state)

$$S = \begin{pmatrix} {}_-\langle\langle C\theta\phi|C\theta\phi\rangle\rangle_+ & {}_-\langle\langle C\theta\phi|D\phi\rangle\rangle_+ & {}_-\langle\langle C\theta\phi|B\theta\rangle\rangle_+ & {}_-\langle\langle C\theta\phi|M_A\rangle\rangle \\ {}_-\langle\langle D\phi|C\theta\phi\rangle\rangle_+ & {}_-\langle\langle D\phi|D\phi\rangle\rangle_+ & {}_-\langle\langle D\phi|B\theta\rangle\rangle_+ & {}_-\langle\langle D\phi|M_A\rangle\rangle \\ {}_-\langle\langle B\theta|C\theta\phi\rangle\rangle_+ & {}_-\langle\langle B\theta|D\phi\rangle\rangle_+ & {}_-\langle\langle B\theta|B\theta\rangle\rangle_+ & {}_-\langle\langle B\theta|M_A\rangle\rangle \\ \langle\langle M_A|C\theta\phi\rangle\rangle_+ & \langle\langle M_A|D\phi\rangle\rangle_+ & \langle\langle M_A|B\theta\rangle\rangle_+ & \langle\langle M_A|M_A\rangle\rangle \end{pmatrix}. \tag{9.8}$$

Let us start with the $(1,1)$ component of S. We wish to calculate

$$\begin{aligned} {}_C\langle\langle E', n', out|E, n, in\rangle\rangle_C & \\ &= \int d\omega\, d\nu\, C^{C'^*}(E', n', \omega, \nu) C^C(E, n, \omega, \nu) \\ &+ \int d\omega\, B^{C'^*}(E', n', \omega) B^C(E, n, \omega) \\ &+ \int d\nu\, D^{C'^*}(E', n', \nu) D^C(E, n, \nu). \end{aligned} \tag{9.9}$$

We rewrite Eq. (9.9) in terms of the lower sector physical states using Eqs. (6.3) to get

$$\int d\omega\, d\nu \left[\int d\lambda'\, \rho'^*_{\lambda',B}(\nu) b^{C'^*}(E', n', \lambda', \omega) + \rho'^*_B(\nu) b^{C'}{}_F{}^*(E', n', M_B, \omega)\right]$$

$$
\begin{aligned}
&\qquad \left[\int d\lambda\, \rho_{\lambda,B}(\nu) b^{C}(E,n,\lambda,\omega) + \rho_B(\nu) b^{C}_{F}(E,n,M_B,\omega)\right] \\
&+ \int d\omega \left[\int d\lambda'\, \sigma'_{\lambda',B} b^{C'^*}(E',n',\lambda',\omega) + \sqrt{Z_B} {b^{C'}_{F}}^{*}(E',n',M_B,\omega)\right] \\
&\qquad \left[\int d\lambda\, \sigma_{\lambda,B} b^{C}(E,n,\lambda,\omega) + \sqrt{Z_B} b^{C}_{F}(E,n,M_B,\omega)\right] \\
&+ \int d\nu\, D^{C'^*}(E',n',\nu) D^{C}(E,n,\nu).
\end{aligned} \tag{9.10}
$$

Doing the integrals over $\lambda$ in Eq. (9.10) we get

$$
\begin{aligned}
&\int d\lambda\, d\lambda'\, \frac{\beta^*(\lambda)}{\beta(\lambda)} \delta(\lambda - \lambda') \int d\omega\, b^{C'^*}(E',n',\lambda',\omega) b^{C}(E,n,\lambda,\omega) \\
&+ \int d\omega\, {b^{C'}_{F}}^{*}(E',n',M_B,\omega) b^{C}_{F}(E,n,M_B,\omega) \\
&+ \int d\nu\, D^{C'^*}(E',n',\nu) D^{C}(E,n,\nu).
\end{aligned} \tag{9.11}
$$

The sum of the first and second integrals gives

$$
\begin{aligned}
&\delta(E-E') \left\{ \delta(n-n') \frac{\beta^*(E-n)\,\alpha^*(n)}{\beta(E-n)\,\alpha(n)} + \frac{2\pi i}{\gamma(E)} \frac{g(E-n)\,f(n)}{\beta(E-n)\,\alpha(n)} \frac{g^*(E-n')\,f^*(n')}{\beta(E-n')\,\alpha(n')} \right\} \\
&- \delta(E'-n'-E+n) \frac{\beta^*(E-n)\,f^*(n')\,f(n)}{\beta(E-n)\,\alpha(n')\,\alpha(n)} \\
&+ \frac{g^*(E'-n')\,f^*(n')\,g(E-n)\,f(n)}{\beta(E'-n')\,\alpha(n')\,\beta(E-n)\,\alpha(n)} \\
&\qquad \left\{ \frac{1}{\gamma(E)\,\alpha(E-E'+n')} + \frac{1}{\gamma(E')\,\alpha(E'-E+n)} \right\} \\
&+ \frac{1}{\gamma(E')\,\gamma(E)} \frac{g^*(E'-n')\,f^*(n')\,g(E-n)\,f(n)}{\beta(E'-n')\,\alpha(n')\,\beta(E-n)\,\alpha(n)} \\
&\quad \left\{ -\int d\lambda\, |\sigma_{\lambda,B}|^2 \frac{1}{\alpha(E'-\lambda)\,\alpha(E-\lambda)} - \frac{Z_B}{\alpha(E'-M_B)\,\alpha(E-M_B)} \right\},
\end{aligned} \tag{9.12}
$$

and the third integral gives

$$
\begin{aligned}
&\delta(E'-n'-E+n) \frac{\beta^*(E-n)\,f^*(n')\,f(n)}{\beta(E-n)\,\alpha(n')\,\alpha(n)} \\
&- \frac{g^*(E'-n')\,f^*(n')\,g(E-n)\,f(n)}{\beta(E'-n')\,\alpha(n')\,\beta(E-n)\,\alpha(n)} \\
&\qquad \left\{ \frac{1}{\gamma(E)\,\alpha(E-E'+n')} + \frac{1}{\gamma(E')\,\alpha(E'-E+n)} \right\}
\end{aligned}
$$

$$
\begin{aligned}
&+\frac{1}{\gamma\left(E'\right)\gamma\left(E\right)}\frac{g^{*}\left(E'-n'\right)f^{*}\left(n'\right)g\left(E-n\right)f\left(n\right)}{\beta\left(E'-n'\right)\alpha\left(n'\right)\beta\left(E-n\right)\alpha\left(n\right)}\\
&\quad\left\{\frac{1}{2}\int d\mu\left|\sigma_{\mu,D}\right|^{2}\frac{1}{\beta\left(E-\mu\right)}\left(\frac{1}{\alpha^{*}\left(E'-E+\mu\right)}+\frac{1}{\alpha\left(E'-E+\mu\right)}\right)\right.\\
&\quad+\frac{1}{2}\int d\mu'\left|\sigma_{\mu',D}\right|^{2}\frac{1}{\beta\left(E'-\mu'\right)}\left(\frac{1}{\alpha^{*}\left(E-E'+\mu'\right)}+\frac{1}{\alpha\left(E-E'+\mu'\right)}\right)\\
&\quad+\frac{Z_{D}}{2\beta\left(E-M_{D}\right)}\left(\frac{1}{\alpha^{*}\left(E'-E+M_{D}\right)}+\frac{1}{\alpha\left(E'-E+M_{D}\right)}\right)\\
&\quad\left.+\frac{Z_{D}}{2\beta\left(E'-M_{D}\right)}\left(\frac{1}{\alpha^{*}\left(E-E'+M_{D}\right)}+\frac{1}{\alpha\left(E-E'+M_{D}\right)}\right)\right\}. \qquad (9.13)
\end{aligned}
$$

Adding Eqs. (9.12) and (9.13), and converting the sum of the integrals to contour integrals (which evaluate to their residues and cancel the other terms with them inside the curly brackets), we are left with

$$
\begin{aligned}
{}_{C}\langle\langle E',n',in|E,n,out\rangle\rangle_{C}=\delta\left(E-E'\right)&\left\{\delta\left(n-n'\right)\frac{\beta^{*}\left(E-n\right)\alpha^{*}\left(n\right)}{\beta\left(E-n\right)\alpha\left(n\right)}\right.\\
&\left.+\frac{2\pi i}{\gamma\left(E\right)}\frac{g\left(E-n\right)f\left(n\right)}{\beta\left(E-n\right)\alpha\left(n\right)}\frac{g^{*}\left(E-n'\right)f^{*}\left(n'\right)}{\beta\left(E-n'\right)\alpha\left(n'\right)}\right\}. \qquad (9.14)
\end{aligned}
$$

In a similar fashion, we can do all the other S-matrix elements. They are

$$
\begin{aligned}
{}_{D}\langle\langle E',out|E,in\rangle\rangle_{D}&=\delta\left(E-E'\right)\left\{\frac{\beta^{*}\left(E-M_{D}\right)}{\beta\left(E-M_{D}\right)}\right.\\
&\qquad\left.+\frac{2\pi iZ_{D}}{\gamma\left(E\right)}\frac{\left|g\left(E-M_{D}\right)\right|^{2}}{\beta\left(E-M_{D}\right)\beta\left(E-M_{D}\right)}\right\}, \qquad (9.15)\\
{}_{B}\langle\langle E',out|E,in\rangle\rangle_{B}&=\delta\left(E-E'\right)\left\{\frac{\alpha^{*}\left(E-M_{B}\right)}{\alpha\left(E-M_{B}\right)}\right.\\
&\qquad\left.+\frac{2\pi iZ_{B}}{\gamma\left(E\right)}\frac{\left|f\left(E-M_{B}\right)\right|^{2}}{\alpha\left(E-M_{B}\right)\alpha\left(E-M_{B}\right)}\right\}, \qquad (9.16)\\
{}_{A}\langle\langle M_{A}|M_{A}\rangle\rangle_{A}&=1, \qquad (9.17)\\
{}_{B}\langle\langle E',out|E,in\rangle\rangle_{D}&=2\pi i\delta\left(E'-E\right)\frac{\sqrt{Z_{D}}\sqrt{Z_{B}}}{\gamma\left(E\right)}\frac{f^{*}\left(E-M_{B}\right)g\left(E-M_{D}\right)}{\alpha\left(E-M_{B}\right)\beta\left(E-M_{D}\right)}, \qquad (9.18)\\
{}_{D}\langle\langle E',out|E,in\rangle\rangle_{B}&=2\pi i\delta\left(E'-E\right)\frac{\sqrt{Z_{D}}\sqrt{Z_{B}}}{\gamma\left(E\right)}\frac{f\left(E-M_{B}\right)g^{*}\left(E-M_{D}\right)}{\alpha\left(E-M_{B}\right)\beta\left(E-M_{D}\right)}, \qquad (9.19)\\
{}_{D}\langle\langle E',out|E,n,in\rangle\rangle_{C}&=2\pi i\delta\left(E'-E\right)\frac{\sqrt{Z_{D}}}{\gamma\left(E\right)}\frac{g\left(n\right)f\left(E-n\right)}{\alpha\left(n\right)\beta\left(E-n\right)}\frac{g^{*}\left(E-M_{D}\right)}{\beta\left(E-M_{D}\right)}, \qquad (9.20)
\end{aligned}
$$

$$
{}_{C}\langle\langle E', n', out|E, in\rangle\rangle_{D} = 2\pi i\delta\left(E' - E\right)\frac{\sqrt{Z_D}}{\gamma\left(E\right)}\frac{g^{*}\left(n'\right)f^{*}\left(E-n'\right)}{\alpha\left(n'\right)\beta\left(E-n'\right)}\frac{g\left(E-M_D\right)}{\beta\left(E-M_D\right)}, \tag{9.21}
$$

$$
{}_{B}\langle\langle E', out|E, n, in\rangle\rangle_{C} = 2\pi i\delta\left(E' - E\right)\frac{\sqrt{Z_B}}{\gamma\left(E\right)}\frac{g\left(n\right)f\left(E-n\right)}{\alpha\left(n\right)\beta\left(E-n\right)}\frac{f^{*}\left(E-M_B\right)}{\alpha\left(E-M_B\right)}, \tag{9.22}
$$

$$
{}_{C}\langle\langle E', n', out|E, in\rangle\rangle_{B} = 2\pi i\delta\left(E' - E\right)\frac{\sqrt{Z_B}}{\gamma\left(E\right)}\frac{g^{*}\left(n'\right)f^{*}\left(E-n'\right)}{\alpha\left(n'\right)\beta\left(E-n'\right)}\frac{f\left(E-M_B\right)}{\alpha\left(E-M_B\right)}, \tag{9.23}
$$

$$
{}_{A}\langle\langle M_A|E, n, in\rangle\rangle_{C} = 0, \tag{9.24}
$$

$$
{}_{C}\langle\langle E', n', out|M_A\rangle\rangle_{A} = 0, \tag{9.25}
$$

$$
{}_{A}\langle\langle M_A|E, in\rangle\rangle_{D} = 0, \tag{9.26}
$$

$$
{}_{D}\langle\langle E', out|M_A\rangle\rangle_{A} = 0, \tag{9.27}
$$

$$
{}_{A}\langle\langle M_A|E, in\rangle\rangle_{B} = 0, \tag{9.28}
$$

$$
{}_{B}\langle\langle E', out|M_A\rangle\rangle_{A} = 0. \tag{9.29}
$$

## 10. Unitarity of the S Matrix

We can almost trivially show that the S matrix that we have obtained is unitary. In equations, we wish to show that

$$
SS^{\dagger} = 1. \tag{10.1}
$$

Let us calculate the $(1,1)$ term in $SS^{\dagger}$. It is

$$
\begin{aligned}
&SS^{\dagger}_{(1,1)} \\
&\quad = \delta\left(E-E'\right)\int dn''\left[\delta\left(n-n''\right)\frac{\beta^{*}\left(E-n\right)\alpha^{*}\left(n\right)}{\beta\left(E-n\right)\alpha\left(n\right)}\right. \\
&\qquad\qquad\qquad\left. + \frac{2\pi i}{\gamma\left(E\right)}\frac{f^{*}\left(n''\right)f\left(n\right)g^{*}\left(E-n''\right)g\left(E-n\right)}{\alpha\left(n''\right)\alpha\left(n\right)\beta\left(E-n''\right)\beta\left(E-n\right)}\right] \\
&\qquad\qquad\left[\delta\left(n'-n''\right)\frac{\beta\left(E-n'\right)\alpha\left(n'\right)}{\beta^{*}\left(E-n'\right)\alpha^{*}\left(n'\right)}\right.
\end{aligned}
$$

$$- \frac{2\pi i}{\gamma^* (E)} \frac{f (n'') f^* (n') g (E - n'') g^* (E - n')}{\alpha^* (n'') \alpha^* (n') \beta^* (E - n'') \beta^* (E - n')} \Bigg]$$

$$- \frac{(2\pi i)^2}{|\gamma (E)|^2} \delta (E - E') \frac{f (n) g (E - n)}{\alpha (n) \beta (E - n)} \frac{f^* (n') g^* (E - n')}{\alpha^* (n') \beta^* (E - n')}$$

$$\left[ Z_D \frac{|g (E - M_D)|^2}{|\beta (E - M_D)|^2} + Z_B \frac{|f (E - M_B)|^2}{|\alpha (E - M_B)|^2} \right]. \tag{10.2}$$

Doing the integral in Eq. (10.2) with the help of Eq. (A.15) we find that the result is $\delta(E - E')\delta(n - n')$, exactly as desired. The rest of the terms are done in the same way. We find

$$SS^\dagger_{(2,2)} = \delta (E - E'), \tag{10.3}$$

$$SS^\dagger_{(3,3)} = \delta (E - E'), \tag{10.4}$$

$$SS^\dagger_{(4,4)} = 1, \tag{10.5}$$

with all other terms in $SS^\dagger$ being zero, as required. Thus $SS^\dagger = \mathbf{1}$. In the same way, we can also show that $S^\dagger S = \mathbf{1}$, and therefore, our S-matrix is unitary.

## 11. Eigenphases of the S Matrix

The interesting case for the S Matrix is when $E > 0$ so that all channels are open. The S Matrix must satisfy

$$S\zeta = \tau\zeta, \tag{11.1}$$

where $|\tau|^2 = 1$, for some $\zeta$. This is equivalent to the following relations (where we ignore the discrete $A$ channel, as it is decoupled from everything else, and suppress $\delta(E - E')$)

$$\tau - \frac{\beta^* (E - n) \alpha (n)}{\beta (E - n) \alpha (n)} \zeta_n$$

$$= \frac{2\pi i}{\gamma(E)} \frac{f(n)g(E-n)}{\alpha(n)\beta(E-n)} \bigg\{ \int dn' \, \frac{f^*(n')g^*(E-n')}{\alpha(n')\beta(E-n')} \zeta_{n'} + \sqrt{Z_D} \frac{g^*(E-M_D)}{\beta(E-M_D)} \zeta_D + \sqrt{Z_B} \frac{f^*(E-M_B)}{\alpha(E-M_B)} \zeta_B \bigg\}, \tag{11.2$a$}$$

$$\tau - \frac{\beta^*(E-M_D)}{\beta(E-M_D)} \zeta_D = \frac{2\pi i}{\gamma(E)} \sqrt{Z_D} \frac{g(E-M_D)}{\beta(E-M_D)} \bigg\{ \int dn' \, \frac{f^*(n')g^*(E-n')}{\alpha(n')\beta(E-n')} \zeta_{n'} + \sqrt{Z_D} \frac{g^*(E-M_D)}{\beta(E-M_D)} \zeta_D + \sqrt{Z_B} \frac{f^*(E-M_B)}{\alpha(E-M_B)} \zeta_B \bigg\}, \tag{11.2$b$}$$

$$\tau - \frac{\alpha^*(E-M_B)}{\alpha(E-M_D)} \zeta_B = \frac{2\pi i}{\gamma(E)} \sqrt{Z_B} \frac{f(E-M_B)}{\alpha(E-M_B)} \bigg\{ \int dn' \, \frac{f^*(n')g^*(E-n')}{\alpha(n')\beta(E-n')} \zeta_{n'} + \sqrt{Z_D} \frac{g^*(E-M_D)}{\beta(E-M_D)} \zeta_D + \sqrt{Z_B} \frac{f^*(E-M_B)}{\alpha(E-M_B)} \zeta_B \bigg\}. \tag{11.2$c$}$$

We now define the unimodular quantities

$$\tau(n) \equiv \frac{\beta^*(E-n)\,\alpha^*(n)}{\beta(E-n)\,\alpha(n)}, \tag{11.3$a$}$$

$$\tau_D \equiv \frac{\beta^*(E-M_D)}{\beta(E-M_D)}, \tag{11.3$b$}$$

$$\tau_B \equiv \frac{\alpha^*(E-M_B)}{\alpha(E-M_B)}. \tag{11.3$c$}$$

These are the basic equations. We can solve them for continuum values or for discrete values of the eigenphase shifts. Let us start with the continuum values. We invert Eq. (11.2a) and put a delta function on the right hand side along with the appropriate normalization. We then multiply both sides of the equation by $\frac{f^*(n)g^*(E-n)}{\alpha(n)\beta(E-n)}$ and integrate over $n$ to get

$$\bigg\{ 1 - \frac{2\pi i}{\gamma(E)} \int \frac{|f(l)|^2 |g(E-l)|^2}{|\alpha(l)|^2 |\beta(E-l)|^2} \frac{\tau(l)}{\tau - \tau(l) + i\epsilon} \bigg\} \int dn' \, \frac{f^*(n')g^*(E-n')}{\alpha(n')\beta(E-n')} \zeta_{n'} = \frac{f^*(\tau)\,g^*(\tau)}{\alpha(\tau)\,\beta(\tau)} + \frac{2\pi i}{\gamma(E)} \int \frac{|f(l)|^2 |g(E-l)|^2}{|\alpha(l)|^2 |\beta(E-l)|^2} \frac{\tau(l)}{\tau - \tau(l) + i\epsilon} \bigg\{ \sqrt{Z_D} \frac{g^*(E-M_D)}{\beta(E-M_D)} \zeta_D + \sqrt{Z_B} \frac{f^*(E-M_B)}{\alpha(E-M_B)} \zeta_B \bigg\}. \tag{11.4}$$

Defining

$$\Sigma \equiv 1 - \frac{2\pi i}{\gamma(E)} Z_D \frac{|g\,(E-M_D)\,|^2}{|\beta\,(E-M_D)\,|^2} \frac{\sigma_D}{\sigma-\sigma_D} - \frac{2\pi i}{\gamma(E)} Z_B \frac{|f\,(E-M_B)\,|^2}{|\alpha\,(E-M_B)\,|^2} \frac{\sigma_B}{\sigma-\sigma_B}, \tag{11.5}$$

we invert Eq. (11.2b) and Eq. (11.2c) to solve for the term in the curly braces on the right hand side of Eq. (11.4), namely, $\sqrt{Z_D}\frac{g^*(E-M_D)}{\beta(E-M_D)}\zeta_D + \sqrt{Z_B}\frac{f^*(E-M_B)}{\alpha(E-M_B)}\zeta_B$, and find

$$\sqrt{Z_D}\frac{g^*(E-M_D)}{\beta(E-M_D)}\zeta_D + \sqrt{Z_B}\frac{f^*(E-M_B)}{\alpha(E-M_B)}\zeta_B = \left(\frac{1}{\Sigma}-1\right)\int dn'\,\frac{f^*(n')g^*(E-n')}{\alpha(n')\beta(E-n')}\zeta_{n'}. \tag{11.6}$$

We now define

$$\chi(\tau) = 1 - \frac{2\pi i}{\gamma\,(E)\,\Sigma}\int dl\,\frac{|f\,(l)\,|^2|g\,(E-l)\,|^2}{|\alpha\,(l)\,|^2|\beta\,(E-l)\,|^2}\frac{\tau\,(l)}{\tau-\tau\,(l)+i\epsilon}, \tag{11.7}$$

and use this to combine Eq. (11.4) and Eq. (11.6), and find

$$\int dn'\,\frac{f^*(n')g^*(E-n')}{\alpha(n')\beta(E-n')}\zeta_{n'} + \sqrt{Z_D}\frac{g^*(E-M_D)}{\beta(E-M_D)}\zeta_D + \sqrt{Z_B}\frac{f^*(E-M_B)}{\alpha(E-M_B)}\zeta_B = \frac{1}{\chi(\tau)\Sigma}\frac{f^*\,(\tau)\,g^*\,(\tau)}{\alpha\,(\tau)\,\beta\,(E-\tau)}. \tag{11.8}$$

Therefore, our continuum solutions are

$$\zeta_n = \sqrt{\tau'}\delta\,(\tau-\tau(n)) + \frac{2\pi i}{\gamma(E)}\frac{f(n)g(E-n)}{\alpha(n)\beta(E-n)}\frac{1}{(\tau-\tau(n)+i\epsilon)\,\chi(\tau)\Sigma}\frac{f^*\,(\tau)\,g^*\,(E-\tau)}{\alpha\,(\tau)\,\beta\,(E-\tau)}, \tag{11.9$a$}$$

$$\zeta_D = \frac{2\pi i}{\gamma(E)}\sqrt{Z_D}\frac{g(E-M_D)}{\beta(E-M_D)}\frac{1}{(\tau-\tau_D+i\epsilon)\,\chi(\tau)\Sigma}\frac{f^*\,(\tau)\,g^*\,(E-\tau)}{\alpha\,(\tau)\,\beta\,(E-\tau)}, \tag{11.9$b$}$$

$$\zeta_B = \frac{2\pi i}{\gamma(E)}\sqrt{Z_B}\frac{f(E-M_B)}{\alpha(E-M_B)}\frac{1}{(\tau-\tau_B+i\epsilon)\,\chi(\tau)\Sigma}\frac{f^*\,(\tau)\,g^*\,(E-\tau)}{\alpha\,(\tau)\,\beta\,(E-\tau)}. \tag{11.9$c$}$$

To investigate the spectrum of $\tau$, we use the method of [7]. We define the following quantities, taking advantage of their being unimodular:

$$e^{2i\theta(n)} \equiv \tau(n), \tag{11.10$a$}$$

$$e^{2i\theta_D} \equiv \tau_D, \tag{11.10b}$$

$$e^{2i\theta_B} \equiv \tau_B, \tag{11.10c}$$

$$e^{2i\delta} \equiv \tau. \tag{11.10d}$$

We then put these definitions in Eqs. (11.9), and see that $\tau(n) = e^{2i\theta(n)}$ ranges continuously along a unit circle in the complex plane from $\theta = \theta(0)$ to $\theta = \theta(E)$.

In addition, these solutions are continuum normalized

$$\begin{aligned}\int dn\, \zeta_n\left(\tau' - i\epsilon\right) \zeta_n\left(\tau + i\epsilon\right)& \\ + \zeta_D\left(\tau' - i\epsilon\right) \zeta_D\left(\tau + i\epsilon\right) + \zeta_B\left(\tau' - i\epsilon\right) \zeta_B\left(\tau + i\epsilon\right)& \\ = \delta\left(\tau' - \tau\right),&\end{aligned} \tag{11.11}$$

and will be complete if there are no discrete zeros of $\chi(\tau)$. If there are, they will have to be included in the completeness identity. We now find the number of discrete zeros of $\chi(\tau)$, that is, the number of discrete eigenphase shifts of our S-Matrix.

We define

$$\begin{aligned}\tau \equiv \frac{1+ix}{1-ix}, \quad \tau\left(n\right) \equiv \frac{1+ix\left(n\right)}{1-ix\left(n\right)}, \\ \tau_D \equiv \frac{1+ix_D}{1-ix_D}, \quad \tau_B \equiv \frac{1+ix_B}{1-ix_B},\end{aligned}$$

put these in Eq. (11.7), and take real and imaginary parts to get

$$\begin{aligned}-\frac{1}{2}\int dl\, \frac{|f\left(l\right)|^2 |g\left(E-l\right)|^2}{|\alpha\left(l\right)|^2 |\beta\left(E-l\right)|^2} \frac{x\left(1+x\left(l\right)\right)}{x - x\left(l\right) + i\epsilon}& \\ = \mathrm{Im}\left(\frac{\gamma\left(E\right)}{2\pi i}\right) - \frac{1}{2}\frac{x\left(1+x_D\right)}{x - x_D} Z_D \frac{|g\left(E-M_D\right)|^2}{|\beta\left(E-M_D\right)|^2}& \\ - \frac{1}{2}\frac{x\left(1+x_B\right)}{x - x_B} Z_B \frac{|f\left(E-M_B\right)|^2}{|\alpha\left(E-M_B\right)|^2},&\end{aligned} \tag{11.12a}$$

$$\begin{aligned}-\frac{1}{2}\int dl\, \frac{|f\left(l\right)|^2 |g\left(E-l\right)|^2}{|\alpha\left(l\right)|^2 |\beta\left(E-l\right)|^2}& \\ = \mathrm{Re}\left(\frac{\gamma\left(E\right)}{2\pi i}\right) - \frac{1}{2} Z_D \frac{|g\left(E-M_D\right)|^2}{|\beta\left(E-M_D\right)|^2} - \frac{1}{2} Z_B \frac{|f\left(E-M_B\right)|^2}{|\alpha\left(E-M_B\right)|^2}.&\end{aligned} \tag{11.12b}$$

We observe that Eq. (11.12b) is an identity, by means of Eq. (A.15). To find the number of zeros of $\chi(\tau)$, we multiply both sides of Eq. (11.12a) by $(x - x_D)(x - x_B)$. We then find that the highest power of $x$ appearing in Eq. (11.12a) is $x^3$, barring any higher powers contributed by the integral. Therefore, there are at least three discrete zeros of $\chi(\tau)$, and thus, at least three discrete solutions which will have to be included in the completeness identity, Eq. (11.11).

## 12. A general formalism for scattering theory

In this section, we describe an approach due to Sudarshan and collaborators [19-25], which takes a very different view of scattering problems, and is quite different in spirit. It is essentially immune to many of the problems that occur in the conventional approaches in the literature. The idea is that one always works with the complete set of eigenstates of the full Hamiltonian, $H$, properly labelled. This set, by definition, is both orthonormal and complete. The matrix made up of these eigenstates is the Möller matrix. This Möller matrix, again by definition, will diagonalize the full Hamiltonian giving the comparison Hamiltonian.

In other words, we have a Hamiltonian for the system, $H$. We wish a physical interpretation of this object as a scattering system. One starts by considering the complete set of states for $H$, which we denote as $\psi_\alpha(E)$, where $E$ is the energy of the eigenstate, and $\alpha$ contains everything else necessary to uniquely specify the state such as spin, channel, etc. Then,

$$H\psi_\alpha(E) = E_\alpha \psi_\alpha(E) \,. \tag{12.1}$$

Form the generalized Möller matrix, $W$, by defining

$$W_{E,\alpha} \equiv \psi_\alpha(E).$$

Therefore,

$$HW = WH_C. \tag{12.2}$$

where the implied integration is of the Stielje's type–*i.e.* we sum over any discrete indices, and integrate over any continuous ones. Because we have assumed that the set of eigenstates of $H$ is complete, this Möller matrix has the property that

$$\Omega\Omega^\dagger = \mathbf{1}, \quad \Omega^\dagger\Omega = \mathbf{1}. \tag{12.3}$$

It is thus isometric and unitary, as long as we ensure that the spectra of $H$ and $H_C$ are the same, and the spectrum multiplicity is properly preserved. The only caveat here is that not all formally hermitian Hamiltonians have a complete set of eigenstates. However, all "reasonable" Hamiltonians will have a such complete set.

Now, one normally wants an interpretation of a scattering system in terms of asymptotic states. The point here is that, to get such an interpretation, we should use $H_C$, not $H_0$. We must set up a correspondence between the set of eigenstates of $H_C$ with $H$, because unlike $H_0$, we are guaranteed that $H$ and $H_C$ are isospectral. Any asymptotic conditions (such as the strong convergence properties of Eqs. (2.26)) should be expressed using $H_C$, not $H_0$. Thus, it is $H_C$, not $H_0$, which is the proper starting point for any perturbative scheme. Furthermore, as indicated by our model, we should endeavour to construct a perturbative scheme to calculate the full states, not the asymptotic ones, because we cannot say, with any confidence, what the appropriate conditions are on the asymptotic states (see the next section for details). In addition, note another advantage of this formalism.

Any, and all, shifts in the thresholds and spectrum of $H$ (such as a mass renormalization) are automatically taken care of by this procedure.

Finally, it is worth remarking that, in general, $W$ and $H_C$ are not going to be analytic in the coupling constants. Therefore, the procedure that one occasionally sees in the literature of splitting $H$ into $H_C + V'$ is not very useful, and not very constructive. Both $H_C$ and $V'$ will be complicated functions of the coupling constants, and will have all sorts of renormalization factors appearing.

If we are interested in working perturbatively, then we must set up the system carefully. We give here an analysis of Sudarshan [26]. Consider a quantum system defined in a Hilbert space $\mathcal{H}$ with the Hamiltonian split in the usual way

$$H = H_0 + V, \tag{12.4}$$

in which we already know the ideal eigenstates for the continuum, and the proper eigenvectors for the discrete states. The ideal states are, of course, not normalizable and we must take proper linear combinations of them to get states that are square integrable, and in $\mathcal{H}$. Then, we set up a correspondence between eigenstates of $H$ and eigenstates of $H_0$, in such a manner that

$$H\psi_\lambda = \lambda\psi_\lambda, \tag{12.5$a$}$$

$$H_0\psi_{0\lambda} = \lambda\psi_{0\lambda}. \tag{12.5$b$}$$

Therefore,

$$\left(1 - G_0\left(\lambda\right) V\right) \psi_\lambda = \psi_{0\lambda}, \tag{12.6$a$}$$

$$G_0\left(\lambda\right) \left(\lambda - H_0\right) = \mathbf{1}, \tag{12.6$b$}$$

where $G_0$ is the free Green's function. From this we can write

$$\psi_\lambda = \left(1 - G_0\left(\lambda\right)V\right)^{-1}\psi_{0\lambda}, \tag{12.7}$$

thus defining for us a possible Möller matrix, $\Omega'$ given by

$$\Omega' = \left(1 - G_0 V\right)^{-1}. \tag{12.8}$$

Now this $\Omega'$ is a possible Möller matrix in the sense that it intertwines $H$ and $H_0$:

$$H\Omega' = \Omega' H_0. \tag{12.9}$$

Unfortunately, it is not very useful because it is not necessarily unitary, or even isometric. One must renormalize it correctly so as to get a unitary operator. Furthermore, $H$ and $H_0$ are not isospectral. Consider the full Green's function, $\mathcal{G}(\lambda)$:

$$\mathcal{G}\left(\lambda\right) = \frac{1}{\lambda - H + i\epsilon} = \left(1 - G_0\left(\lambda\right)V\right)^{-1} G_0\left(\lambda\right). \tag{12.10}$$

While, at first glance, it would seem that $G_0$ and $\mathcal{G}$ have the same singularities, this is not necessarily true. Firstly, $\mathcal{G}$ can have additional singularities from the first factor $(1 - G_0(\lambda)V)^{-1}$ in Eq. (12.10). These can come from bound states produced by the interaction, and more importantly, from continuum states in which one or more of the particles is composite so that its mass gets shifted. Secondly, $\mathcal{G}$ can have some of its singularities cancelled when this factor vanishes. Therefore, $\mathcal{G}$ and $G_0$ are not necessarily isospectral, in general. In other words, the statement that "perturbations vanish at infinity" is not valid generally. Rather, this näive asymptotic condition is not generally fulfilled. This shows us why $\Omega'$ failed to be unitary: the new spectra produced by $(1 - G_0(\lambda)V)^{-1}$ do not appear with a canonical weight. As advertised, we correct this problem by defining a renormalized

$\Omega$ given by

$$\Omega = (1 - G_0 V)^{-1} D^{-1}, \tag{12.11a}$$

where

$$D^2 = \left(1 - V G_0^{+}\right)^{-1} (1 - G_0 V)^{-1}. \tag{12.11b}$$

However, since this new $\Omega$ is unitary, it connects $H$ with an isospectral and diagonal Hamiltonian. We have already seen that this associated diagonal Hamiltonian cannot be $H_0$. Rather, it is a different object, which we call the comparison Hamiltonian, $H_C$. For further details, and concrete examples of this formalism applied to several models, such as the Lee Model, the separable potential model, and the Cascade model, see [26]. In these models, one can explicitly see these various effects such as shifts in the continuous spectra, the deletion of spectra from $H_0$ to get the spectra of $H$, and the augmentation of spectra in $H_0$ to get the spectra of $H$.

## 13. Putting the "generic" formalisms to the test

There are many different approaches to Quantum Scattering in the literature. The most familiar of these is potential scattering. Others include the LSZ formalism, the "almost local" formalism, and the Lax-Phillips formalism. Lax-Phillips [27] theory is outside the scope of this work.

The well known LSZ formalism [5], extended by Mohan [28], postulates the convergence of the matrix elements of interacting fields to the matrix elements of free fields. However, the formalism does not apply in many cases. For example, as noted by LSZ themselves, it is inapplicable to problems in which stable bound states exist. Trouble occurs when this point is forgotten, and the formalism is extended into areas where it is inapplicable. The "almost local" formalism due to Haag [1], Ruelle [9], Ekstein [10], Jauch [12], Araki [29], and others tries to be general enough to consider complicated problems [1]. Its basic idea is that it is possible to construct asymptotic in-going and out-going states as strong limits in Hilbert space, if a certain "space like asymptotic condition" is verified by the vacuum expectation values of products of field operators [9]: the so called "almost local" operators [1].

We shall restrict our attention to the conventional, and quite "generic" formalism, as reviewed earlier in sections 2 and 3; as mentioned before, the LSZ formalism is *not applicable* to situations where stable bound states are present, such as our model. We will compare these results to the results obtained from the Rearrangement Model.

Conventional formalisms for quantum scattering theory have the following protocol for generic scattering systems:

1. They do not use the comparison Hamiltonian
2. The asymptotic states are orthonormal [1,9,13]
3. The completeness of the asymptotic states is postulated [1,13]
4. For the case of potential scattering only, the Möller Matrix is isometric but not necessarily unitary [1,13]
5. The eigenstates of the exact Hamiltonian are never considered.

We shall take up these points one by one, and put them to the test by comparing them to the results explicitly obtained from our model.

1. It is essential when taking the limits, $\lim_{t\to\pm\infty} e^{iHt}e^{-iH_0t}\Psi$, where $\Psi$ is either a wave function or a field operator, that the continuous spectra of $H$ and $H_0$ coincide. If they did not there would be wild oscillations while taking the limit, and the limit would not exist. It is for this purpose that $H_0$ is mass-renormalized to $H_0'$. However, in general, this is still not enough. It is perfectly possible, if there are bound states or unstable particles in the spectrum of $H$, that no amount of tinkering with $H_0$ will make its spectrum coincide with $H$. This can be seen by inspection of Eqs. (8.7) and (8.8). No amount of renormalization of $H_0$ can give us the discrete $M_A$ state present in $H_C$, but this may be ignored because $M_A$ is a discrete point eigenvalue. On the other hand, we do have the possibility of a continuous spectrum in $H$ corresponding to the scattering states involving physical $B$ or $D$ particles.

However, unlike $H$ and $H_0$, $H$ and $H_C$ are guaranteed to be isospectral because $H_C$ is obtained by diagonalizing $H$. Therefore, it is $H_C$, and not $H_0$, that is the proper starting point for any scattering scheme, perturbative or otherwise. The method for obtaining the correct spectrum of $H$ by perturbation theory is discussed in the work of Sudarshan, Chiu, and Bhamathi [30]. In simple cases such as when stable bound states are not present, or field theory with no bound states or unstable particles, $H_C$ can be identified with the renormalized $H_0$, as noted in section 8.

In fact, even in cases where (formally) no splitting is made, *i.e.* no explicit mention or use is made of an $H_0$, there is still the implicit use of $H_0$ because, commonly, asymptotic particles are defined as solutions of free particle equations like the Klein-Gordon equation.

2. Both formalisms assert the orthonormality of the asymptotic states, and the result is supposed to be generic. In Eqs. (9.4), we have obtained the asymptotic states of the Rearrangement Model according to both formalisms. Yet, as we can see from a glance at the Haag-Ruelle asymptotic wave functions, Eqs. (9.4), the asymptotic states computed according to their rules do not form an orthonormal set. This point should not cause confusion. Our full states, namely, Eqs. (6.4), (6.5), (6.6), and (6.7) are, indeed, all orthonormal to each other, as was shown in section VII. As a result, we have orthonormal sets of "in" and "out" states. However, when we calculate the Haag-Ruelle type asymptotic states according to either of the formalisms, we find that they are not orthonormal. This lack of orthonormality stems from a factor of the wave function renormalization constant that appears in each of the asymptotic wave functions. This factor is essential: if it were not present, the interacting states would not be orthonormal.

3. As mentioned earlier, Ruelle extends Haag's work by postulating the completeness of the "in" and "out" states [9]. This is also postulated in simple potential scattering [13]. This postulate is necessary to prove that the S-matrix is unitary. Again, simply by inspection of Eqs. (9.4), we can see that the asymptotic states of the Rearrangement Model, according to these two formalisms, are not complete. Again, this point should not cause confusion. Our full states, Eqs. (6.4), (6.5), (6.6), and (6.7) are complete, as was shown in section VII.

As a result, our "in" and "out" states form complete sets. However, the set of Haag-Ruelle type asymptotic states calculated according to either of the two formalisms is *not* complete.

4. In potential scattering the Möller matrix, $\Omega$, can be defined using the full interacting wave functions so that it is isometric even in the presence of bound states [1]. We see that the Haag-Ruelle asymptotic solutions, Eqs. (9.4), obtained by the use of $\Omega$, are certainly not orthonormal, whereas the original interacting wave functions were; therefore, the Möller matrix computed by their rules is not isometric, *i.e.* it is not norm preserving. However, the generalized Möller matrix that we defined in Eq. (8.1) is not only isometric, but unitary.

5. It is important to note that even though these asymptotic wave functions are neither orthonormal nor complete, they still lead to the correct S-matrix, as can be verified by calculating it using Eq. (9.6). If we had insisted upon the asymptotic wave functions being orthonormal and complete, we would have got the wrong S-matrix.

6. Notice that because of this lack of orthonormality and completeness in the exact asymptotic states, the strong limits of Eqs. (2.26) are satisfied. Namely,

$$\lim_{t\to-\infty}\left[\Psi(t)-\psi_{\text{in}}(t)\right]\Rightarrow 0, \tag{13.1$a$}$$

$$\lim_{t\to+\infty}\left[\Psi(t)-\psi_{\text{out}}(t)\right]\Rightarrow 0, \tag{13.1$b$}$$

$$\lim_{t\to-\infty}\psi_{\text{in}}(0)\Rightarrow\Psi(0)\equiv\Omega^{(+)}\psi_{\text{in}}(0), \tag{13.1$c$}$$

are automatically satisfied. This can be seen easily in the following way. For the first two equations above, the expression on the left hand side is just the requisite full wave function, but with the delta function part removed. When we now take the norm and then take the limit, the remainder cancels giving zero. Similarly, the third equation above can be shown to be satisfied.

On the other hand, if we had insisted that the asymptotic states *are* orthonormal and complete, the wave function renormalization constants would have been missing from the asymptotic states. Thus, the delta function pieces would not have cancelled between the full and asymptotic states, and therefore, these pieces would contribute, and we would get a non-zero result. Thus, in this case, the strong limit would not hold.

7. It is important to note that the reason that all these problems occur is that the full states are never considered. Most formalisms in the literature try to set up the problem in terms of the asymptotic states, and are thus forced to make assumptions regarding their properties and behavior. These assumptions are not necessarily correct in general, as is amply demonstrated by the Rearrangement Model, and other models such as the Cascade Model [7].

All this points out the importance of the correct normalization of the state vectors, a point already considered by DeWitt [31]. However, his work was restricted to the case of no bound states. The question of the correct description of the asymptotic states was also considered by Van Hove in his papers on the description of "persistent interactions" [4]. However, as noted in those papers, the formalism developed there does not deal with cases involving bound states, and does not deal with field theoretic scattering except for a few comments at the end.

In the multichannel case (such as rearrangement collisions), in the "channel Hamiltonian" formalism, the statement is made that the basis states of one group of channels are not orthogonal to the others [10,11,13] because they are eigenstates of different free Hamiltonians. As we can see, in our model, the physical states $C\theta\phi$, $D\phi$, and $B\theta$ are strictly orthogonal to each other. Evidently, this problem arises due to the use of "channel Hamiltonians" in the formalism. It is our belief that the method of splitting up the interaction differently depending on which channel one is considering is fundamentally flawed because "every channel can be distinguished and is observable independently in experiments. This means that these channels should be orthogonal to each other" [32]. One method for ensuring orthonormality is given in [32]; however, this method still suffers from the flaws pointed out above.

It is straightforward to see the problems caused by this lack of orthogonality. We are instructed, in these formalisms, to begin with asymptotic states. Let us first consider the "channel Hamiltonian" formalism. Then, the asymptotic states are the eigenstates of the "channel Hamiltonian" in the sector we are considering. As an example, let us consider

$$|M_B\theta(\omega)\rangle\rangle \to |M_D\phi(\nu)\rangle\rangle, \tag{13.2}$$

where $M_B$ is the physical $B$ particle and $M_D$ is the physical $D$ particle. We immediately notice, even before we consider any scattering, that the $|M_B\theta(\omega)\rangle\rangle$ state is not orthogonal to the $|M_D\phi(\nu)\rangle\rangle$ state, as can be seen by inspection of Eqs. (5.10), (5.11), (5.12), (5.13), and (5.14). In other words, two experimentally distinct channels are not orthogonal to each other. This will clearly lead to the wrong S-matrix elements because it says that even if there is no scattering, there is a non-zero probability that the $|M_B\theta(\omega)\rangle\rangle$ state will turn into the $|M_D\phi(\nu)\rangle\rangle$ state. We cannot even argue that the two states are "asymptotically orthonormal" [10] because they clearly are not. This can easily be seen by observing that both $|M_B\theta(\omega)\rangle\rangle$ and $|M_D\phi(\nu)\rangle\rangle$ have expansion coefficients in the "bare" $|C\theta(\omega)\phi(\nu)\rangle$ sector. Therefore, as these states are neither orthonormal nor complete, we cannot have an isometric or unitary S-matrix, since orthonormality is necessary for isometry, and completeness for unitarity. However, we have constructed a set of orthonormal (and complete) solutions for our system, a feat that many authors [33] tacitly assume is not possible, and have a perfectly isometric and unitary S-matrix.

These problems with the S-matrix can be verified by explicit calculation. Since the calculation is tedious, we describe the method, and leave it to the interested reader to verify the results. Our interest is in the scattering of physical states, and so we must start by re-expressing the Hamiltonian, Eqs. (4.3) and (4.4), in terms of the operators which create the *physical* $B$ and $D$ particles. We denote these operators by $\mathcal{B}$ and $\mathcal{D}$, respectively. They are found by inspection of Eqs. (5.10), (5.11), (5.12), (5.13), and (5.14), which are the wave functions for the physical particles. To find the expressions for these operators, we promote the states $|C\phi(\nu)\rangle$, $|C\theta(\omega)\rangle$, $|B\rangle$, and $|D\rangle$ to operators, all acting on the vacuum, and read off the expansions for the operators $\mathcal{B}$ and $\mathcal{D}$. In other words,

$$\mathcal{B}^\dagger = \int d\nu\, \rho_B(\nu)\, C^\dagger \phi^\dagger(\nu) + \sqrt{Z_B} B^\dagger, \tag{13.3a}$$

$$\mathcal{D}^\dagger = \int d\omega\, \rho_D(\omega)\, C^\dagger \theta^\dagger(\omega) + \sqrt{Z_D} D^\dagger. \tag{13.3b}$$

We re-express the Hamiltonian in terms of these operators, which can be split into various channel Hamiltonians, from which the S-matrix is calculated.

We can go even further than this. Consider the state $|M_B\theta(\omega)\rangle\rangle$, which is a product state of the physical $B$ particle and a free $\theta$ particle. If we wanted the asymptotic state corresponding to this then, by the Haag-Ruelle protocol, we should find that the components of this state are only in the $|C\theta\phi\rangle$ and $|B\theta\rangle$ sectors, with no admixture of the $|D\phi\rangle$ state. However, we can use our exact solutions to calculate this asymptotic state. We will find that this assertion will not hold true.

To calculate the asymptotic state, we take the limit

$$\lim_{t\to-\infty} e^{iH_C t} e^{-iHt} |M_B\theta\,(\omega)\,\rangle\rangle. \tag{13.4}$$

Inserting a complete set of states, we have

$$\begin{aligned} &\int dE\, dn \lim_{t\to-\infty} e^{iH_C t} e^{-iHt} |E,n\rangle\rangle\langle\langle E,n|M_B\theta\,(\omega)\,\rangle\rangle \\ &+ \int dE\, dn \lim_{t\to-\infty} e^{iH_C t} e^{-iHt} |E\rangle\rangle_{D\,D}\langle\langle E|M_B\theta\,(\omega)\,\rangle\rangle \\ &+ \int dE\, dn \lim_{t\to-\infty} e^{iH_C t} e^{-iHt} |E\rangle\rangle_{B\,B}\langle\langle E|M_B\theta\,(\omega)\,\rangle\rangle. \end{aligned} \tag{13.5}$$

Expanding each of the physical states, $|E,n\rangle\rangle$, $|E\rangle\rangle_D$, and $|E\rangle\rangle_B$ in terms of the bare states $|C\theta(\omega)\phi(\nu)\rangle$, $|B\theta(\omega)\rangle$, and $|D\phi(\nu)\rangle$, and taking the limit, it is immediately obvious that the expansion coefficients in the $|D\phi(\nu)\rangle$ sector are not zero.

As a physical example, consider the case of a proton bound to a fixed nucleus by a potential $V_P$, and bombarded by a neutron which interacts with the proton and the nucleus through the potentials $V_{PN}$ and $V_N$, respectively [32]. The total Hamiltonian of the system is

$$H = K_P + K_N + V_P + V_N + V_{PN}, \tag{13.6}$$

where $K_P$ and $K_N$ are the kinetic energy of the proton and the neutron, respectively. The initial state, denoted by $\Phi_{1,i}$, is given by

$$H_1\Phi_{1,i} = E_i\Phi_{1,i}, \tag{13.7}$$

where

$$H = H_1 + V_1, \tag{13.8a}$$

$$H_1 = K_P + K_N + V_P, \tag{13.8b}$$

$$V_1 = V_{PN} + V_N. \tag{13.8c}$$

Therefore, the initial state, $\Phi_{1,i}$, is a product of a bound proton, $\phi_P^B(E_i^B)$, and of a free neutron (represented by a plane wave), $u_N(E_i - E_i^B)$, where $E_i^B$ is the binding energy of the proton.

Several possible reactions can occur giving rise to different final products. Let us consider four such reactions.

1. Elastic or inelastic collisions.

   The proton remains bound to the nucleus, and the neutron is free after the collision. Therefore, the Hamiltonian is divided in the same manner as above.

2. Exchange scattering.

   The neutron knocks out the bound proton and becomes bound to the nucleus. The Hamiltonian is then divided as:

$$H = H_2 + V_2, \tag{13.9a}$$

$$H_2 = K_P + K_N + V_N, \tag{13.9b}$$

$$V_2 = V_{PN} + V_P. \tag{13.9c}$$

   Therefore, the final state is

$$H_2\Phi_{2,f} = E_f\Phi_{2,f}, \tag{13.10a}$$

$$\Phi_{2,f} = u_P\left(E_f - E_f^B\right)\phi_N^B\left(E_f^B\right). \tag{13.10b}$$

3. Ionization.

The neutron knocks out the bound proton and both are free after the collision. The Hamiltonian is then divided as:

$$H = H_3 + V_3, \tag{13.11a}$$

$$H_3 = K_P + K_N, \tag{13.11b}$$

$$V_3 = V_{PN} + V_P + V_N. \tag{13.11c}$$

Therefore, the final state is

$$H_3 \Phi_{3,f} = E_f \Phi_{3,f}, \tag{13.12a}$$

$$\Phi_{3,f} = u_P \left(E_f^P\right) u_N \left(E_f - E_f^P\right). \tag{13.12b}$$

4. Pickup.

The proton and the neutron become bound and form a deuteron. The Hamiltonian is then divided as:

$$H = H_4 + V_4, \tag{13.13a}$$

$$H_4 = K_P + K_N + V_{PN}, \tag{13.13b}$$

$$V_4 = V_P + V_N. \tag{13.13c}$$

Therefore, the final state is

$$H_4 \Phi_{4,f} = E_f \Phi_{4,f}, \tag{13.14a}$$

$$\Phi_{4,f} = u_c \left(X, E_f - E_f^B\right) \phi_{PN}^B \left(r, E_f^B\right). \tag{13.14b}$$

Here, $X$ is the center of mass coordinate of the deuteron, and $r$ is the internal coordinate of the deuteron.

| Property | Conventional Formalism | Rearr. Model | |
|---|---|---|---|
| Asymptotic states normalized? | Yes | No | |
| Asymptotic states orthogonal? | Yes | Yes | |
| Asymptotic states complete? | Yes | No | |
| $\Omega$ isometric? | Yes | No | |
| S-matrix unitary? | Yes | Yes | |
| Strong limit satisfied? | No | Yes | |
| $H_C$ used? | No | Yes | |
| Additional property for the multiple-channel case | | | |
| Phys. states orthog.? | No | No | Yes |

**Table 13.1:** Comparison of the properties of the Rearrangement Model to various scattering formalisms.

The final states given by Eqs. (13.10), (13.12), and (13.14), are eigenstates of different free Hamiltonians. Thus, in general, they are not orthogonal to each other, and the concomitant problems follow.

The reason that these methods do not work properly is that the basis used is one in which bound-state eigenfunctions of the Hamiltonians that bind each fragment are multiplied by plane waves for the fragment motion [13]. In our model, because we have made no breakup, we get the physically reasonable result that the wave functions of the bound states are always orthogonal to the scattering states, and that the basis states of different channels are explicitly orthogonal to each other. We do not have to worry about making the explicit assumption that as the separation between the fragments goes to infinity, the overlap becomes negligible. This assumption may or may not be true, and leads to the problems with "persistent interactions" considered by Van Hove [4].

We compare the results from the conventional formalism with those from the Rearrangement Model in Table 13.1.

In addition, even when it is not stated explicitly in the literature, it is often assumed that the spectra of the bound states and the scattering (continuum) states do not overlap. However, it is possible to construct models in which the spectra of one or more bound states overlap with the continuum [34,35]. Therefore, this assumption is not necessarily true, and will in general depend upon the details of the model under consideration. It is also possible to construct two different potentials which can lead to the same S-matrix with, in one case, redundant poles unnecessary for completeness, and in the other case, with the same poles being absolutely necessary for completeness [36,37]. This points out the need for resisting the temptation to identify the poles of the S-matrix with physical bound states of the system.

More importantly, no authors have as yet worried about the evident normalization problem with the asymptotic states because they are always assumed to be normalized. These states are not normalized in the Rearrangement Model, and consequently, assuming orthonormality of the asymptotic states, in general, is very dangerous. In addition, we notice that in this model even though the asymptotic states are not normalized, the interacting states are.

One approach that tries to avoid all these problems, especially in the cases of unstable particles and bound states, is that of analytic continuation [7,19,23,24,38,39] of the state space $\mathcal{H}$ into a generalized vector space $\mathcal{G}$. This has already been done for the case of the Lee model by Sudarshan, Chiu, and Gorini [19], Parravicini, Gorini, and Sudarshan [20], and by Böhm [40]. For instance, with this method, one can identify resonances and redundant poles, and study the decay of a metastable quantum system. It can also be used for many other things, such as studying the Khalfin observation that the decay of a metastable system with an energy spectrum bounded from below can never be strictly exponential [41]. See the above references for details.

## 14. Summary and conclusion

In this work, we constructed a model that allows rearrangement collisions. We explored the spectra and the complete set of orthonormal (ideal) eigenfunctions of this Rearrangement Model in the Rearrangement Sector. Because of the structure of the effective Hamiltonian in this sector, we were able to solve the model exactly. In a similar fashion as for the Cascade Model [7], we find that the spectra can be interpreted as a $B$ particle with energy $M_B < 0$ coupled to a $\theta$ particle with energy $\omega$, $0 < \omega < \infty$; a $D$ particle with energy $M_D < 0$ coupled to a $\theta$ particle with energy $\nu$, $0 < \nu < \infty$; and a $C$ particle of energy 0 coupled to $\theta$ and $\phi$ particles with energies $\omega$ and $\nu$, $0 < \omega, \nu < \infty$. We see that the interacting field theory has a particle interpretation.

Both the $B$ and the $D$ particles suffer mass renormalizations, and these mass renormalizations alter the threshold of the $B\theta$ and $D\phi$ continua, respectively. In Eqs. (6.5b) and (6.6c), we also see the presence of both the mass and wave function renormalizations of the $B$ and $D$ particles in the plane wave parts of their respective wave functions.

We have throughout emphasized the importance of using the comparison Hamiltonian (the diagonalized form of the effective Hamiltonian) because it is isospectral with the full Hamiltonian. Its spectrum differs from that of the free Hamiltonian by the alteration of the $B\theta$ and $D\phi$ continua, and by the addition of a discrete $A$ state. These effects are non-perturbative and, as emphasized in [7], can only be handled by a renormalized perturbation scheme in which $H_C$, not $H_0$, is taken as the starting point.

Our results are surprising when compared to what we would expect from conventional scattering theory. We find that while the interacting state vectors are normalized, the asymptotic states are not. Moreover, the asymptotic states are neither orthonormal nor complete because of the presence of the wave function renormalization factors in the physical $D\phi$ and $B\theta$ sectors. We note that this lack of orthonormality and completeness is absolutely necessary. If we construct the S-matrix from these states, we get the correct result (*i.e.* it is the same S-matrix as the one constructed from the full state). On the other hand, if we didn't allow the wave function renormalization factors because of our demand that the asymptotic states be orthonormal and complete, we would get the wrong S-matrix. Furthermore, for the strong limits Eqs. (2.26) to hold, we must again make sure to have these non-orthonormal and non-complete states. We also find that our physical $C\theta\phi$, $D\phi$, and $B\theta$ states, while being the basis states for different channels, are strictly orthogonal to each other. Further, the Möller matrix, as defined in the literature is not isometric: it does not preserve the norm of the states. However, we defined a generalized Möller matrix which is not only isometric, but unitary. All these results are contrary to the usual formalisms of quantum scattering theory.

More generally, we argued that the correct procedure, for any Hamiltonian, $H$, is to take its complete set of eigenstates, and an associated isospectral comparison Hamiltonian, $H_C$. The matrix of normalized eigenfunctions of $H$ constitutes the generalized Möller matrix, which is unitary and intertwines $H$ and $H_C$.

This model is a very simple one. However, even this simple model is enough to show the problems with conventional perturbation theory, and the conventional formulations of scattering theory. It is clearly necessary in the light of this model, and previous work on the existence of redundant poles in the scattering amplitude [36,37] and the presence of discrete solutions degenerate in energy with the scattering continuum [34,35], that a fundamental re-examination be made of some of the postulates and assumptions of conventional quantum scattering theory.

## Acknowledgments

A portion of this work first appeared in the Ph.D. Dissertation of one of the authors (S.V.) [42]. This research was supported in part by DOE Grant No. DE-FG03-93ER40757.

## A. Some Useful Formulae

The following formulae are very useful for the calculations in the main text. By our definitions in section 6 we have the following ranges for our variables:

$$0 \leq \lambda \leq \infty, \tag{A.1}$$

$$0 \leq \mu \leq \infty, \tag{A.2}$$

$$0 \leq n \leq \infty, \tag{A.3}$$

with $E$ being free to run over all values.

We then have the easily proved identities

$$|g(\lambda)|^2 = \frac{1}{2\pi i}\left[\beta(\lambda) - \beta^*(\lambda)\right], \tag{A.4}$$

$$|f(\mu)|^2 = \frac{1}{2\pi i}\left[\alpha(\mu) - \alpha^*(\mu)\right], \tag{A.5}$$

$$\frac{|g(\lambda)|^2}{|\beta(\lambda)|^2} = -\frac{1}{2\pi i}\left[\frac{1}{\beta(\lambda)} - \frac{1}{\beta^*(\lambda)}\right], \tag{A.6}$$

$$\frac{|f(\mu)|^2}{|\alpha(\mu)|^2} = -\frac{1}{2\pi i}\left[\frac{1}{\alpha(\mu)} - \frac{1}{\alpha^*(\mu)}\right], \tag{A.7}$$

$$\frac{|g(E-\lambda)|^2}{|\beta(E-\lambda)|^2} = -\frac{1}{2\pi i}\left[\frac{1}{\beta(E-\lambda)} - \frac{1}{\beta^*(E-\lambda)}\right] - Z_B\delta(E-\lambda-M_B), \tag{A.8}$$

$$\frac{|f(E-\mu)|^2}{|\alpha(E-\mu)|^2} = -\frac{1}{2\pi i}\left[\frac{1}{\alpha(E-\mu)} - \frac{1}{\alpha^*(E-\mu)}\right] - Z_D\delta(E-\mu-M_D). \tag{A.9}$$

The equations (A.8) and (A.9) follow because $E-\lambda$ and $E-\mu$ can be less than zero, and thus pick up singularities at $M_B < 0$ and $M_D < 0$, respectively. On the other hand $\lambda$ and $\mu$ are always greater than or equal to zero, and so cannot pick up any singularities.

Another useful identity is

$$\gamma(E) = \int d\lambda\, \frac{|g(\lambda)|^2}{|\beta(\lambda)|^2}\frac{1}{\alpha(E-\lambda)} + \frac{Z_B}{\alpha(E-M_B)} \tag{A.10a}$$

$$= \int d\mu\, \frac{|f(\mu)|^2}{|\alpha(\mu)|^2}\frac{1}{\beta(E-\mu)} + \frac{Z_D}{\beta(E-M_D)}. \tag{A.10b}$$

We can easily show this by means of the contours in Figures 1 and 2. If we convert the integral in Eq. (A.10a) into a contour integral by using Eq. (A.6), we get

$$\left(-\frac{1}{2\pi i}\right)\int_{C_1} dz\, \frac{1}{\beta(z)\,\alpha(E-z)} + \frac{Z_B}{\alpha(E-M_B)}, \tag{A.11}$$

with the contour shown in Figure 1. Then, we make a change of variables from $z$ to $E-z$ to get

$$\left(-\frac{1}{2\pi i}\right)(-1)\int_{C_4} dz\, \frac{1}{\alpha(z)\,\beta(E-z)} + \frac{Z_B}{\alpha(E-M_B)}, \tag{A.12}$$

with the contour shown in Figure 2. Now we deform the contour $C_4$ and write it as the contour $C_3$ plus the circle at infinity, while picking up the contributions from the residues of the integrand. Note that the circle at infinity gives no result, so we have

$$\left(-\frac{1}{2\pi i}\right)(-1)(-1)\int_{C_3} dz\, \frac{1}{\alpha(z)\,\beta(E-z)} + \left(-\frac{1}{2\pi i}\right)(-1)(2\pi i)\frac{Z_D}{\alpha(E-M_D)}$$

$$+\left(-\frac{1}{2\pi i}\right)(-1)(-1)(2\pi i)\frac{Z_B}{\alpha(E-M_B)}+\frac{Z_B}{\alpha(E-M_B)}. \qquad (A.13)$$

The $\sqrt{Z_B}$ terms cancel, and the first two terms are Eq. (A.10b), by definition. Therefore, Eq. (A.10a) is equal to Eq. (A.10b), and the identity is established.

We can similarly show that

$$\begin{aligned}
&\int d\lambda \frac{|g(\lambda)|^2}{|\beta(\lambda)|^2}\frac{1}{\alpha(E-\lambda)}\frac{1}{(\lambda-\nu+i\epsilon)} \\
&\quad = \int d\mu \frac{|f(\mu)|^2}{|\alpha(\mu)|^2}\frac{1}{\beta(E-\mu)}\frac{1}{(E-\mu-\nu+i\epsilon)} - \frac{1}{\alpha(E-\nu)\beta(\nu)} \\
&\quad + \frac{Z_D}{\beta(E-M_D)(E-M_D-\nu+i\epsilon)} - \frac{Z_B}{\alpha(E-M_B)(E-M_B-\omega+i\epsilon)}. \qquad (A.14)
\end{aligned}$$

Using Eqs. (A.7), (A.8), and (A.10) we can get another useful formula

$$\begin{aligned}
&\int dn \frac{|f(n)|^2|g(E-n)|^2}{|\alpha(n)|^2|\beta(E-n)|^2} \\
&\quad = \frac{\gamma(E)-\gamma^*(E)}{(-2\pi i)} - Z_D\frac{|g(E-M_D)|^2}{|\beta(E-M_D)|^2} - Z_B\frac{|f(E-M_B)|^2}{|\alpha(E-M_B)|^2}. \qquad (A.15)
\end{aligned}$$

Finally, in Figure 3, we display the branch cuts and poles of $\frac{1}{\gamma(z)}$ which are used in showing the completeness of our solution set.

## Figure Captions

Figure 1. Contour for Eq. (A.10a)

Figure 2. Contour for Eq. (A.10b)

Figure 3. Contour for the function $\frac{1}{\gamma(z)}$.